\documentclass[aps,pre,preprint,groupedaddress,showpacs,nofootinbib]{revtex4-1}

\usepackage{graphics}
\usepackage{latexsym}
\usepackage{bm}
\usepackage{fancybox}
\usepackage{accents}
\usepackage{mathrsfs}
\usepackage{mathrsfs}
\usepackage{color}
\usepackage{ulem}

\newcommand{\grad}{\nabla}

\newcommand{\wz}{\omega}

\newcommand{\hwz}{\hat{\omega}}

\newcommand{\dwz}{\delta\omega}
\renewcommand{\vr}{\vec{r}}
\newcommand{\vk}{\vec{k}}

\newcommand{\vq}{\vec{q}}
\renewcommand{\div}{\nabla \cdot}
\renewcommand{\u}{\vec{u}}

\renewcommand{\vec}[1]{\bm{#1}}

\begin{document}


\title{Explicit Formula of Energy-Conserving Fokker-Planck Type
Collision Term for Single Species Point Vortex Systems with
Weak Mean Flow}


\author{Yuichi Yatsuyanagi}
\affiliation{Faculty of Education, Shizuoka University, Suruga-ku, Shizuoka 422-8529, Japan}
\author{Tadatsugu Hatori}
\affiliation{National Institute for Fusion Science, Toki, Gifu 509-5292, Japan}
\author{Pierre-Henri Chavanis}
\affiliation{Laboratoire de Physique Th\'eorique (IRSAMC), CNRS and UPS,
Universit\'e de Toulouse, F-31062 Toulouse, France}

\date{\today}

\begin{abstract}
This paper derives a kinetic equation for  a two-dimensional
single species point vortex system.  We
consider a situation (different from the ones considered
previously) 
of weak mean flow where the time scale of the
macroscopic motion is longer than the decorrelation
time so that the
trajectory of the point vortices
can be approximated by a straight line on the decorrelation
time scale. This may be the case when the number $N$ of point
vortices is not too large.
Using a kinetic theory based on the Klimontovich formalism, we
derive
a collision term consisting of a diffusion term and a drift
term, whose structure
is similar to the Fokker-Planck equation.
The collision term exhibits several important properties:
(a) it includes a nonlocal effect;
(b) it conserves the mean field energy; 
(c) it satisfies the $H$ theorem;
(d) its effect vanishes in each local equilibrium region with the same temperature.
When the system reaches a global equilibrium state, the collision term completely
converges to zero all over the system.
The theoretical prediction of the relaxation 
time of the system from the obtained kinetic equation is confirmed by direct
numerical simulations of point vortices.
\end{abstract}
\pacs{47.32.C-, 47.27.tb, 05.10.Gg, 05.20.Dd, 52.27.Jt, 47.27.-i}


\maketitle

\section{Introduction}
The two-dimensional (2D) microscopic point vortex system
\cite{Newton} is a
formal solution
of the 2D inviscid microscopic Euler equation, 
\begin{equation}
	\frac{\partial }{\partial t} \hwz(\vr,t) + \div (\hat{\u}(\vr,t)
\hwz(\vr,t)) = 0, \label{eqn:euler}
\end{equation}
where $\hwz(\vr,t)$ and $\hat{\u}(\vr,t)$ are the microscopic vorticity and the microscopic
velocity, respectively.
Equation (\ref{eqn:euler}) is formally identical to the
macroscopic
Euler equation
\begin{equation}
	\frac{\partial }{\partial t} \wz(\vr,t) + \div (\u(\vr,t) \wz(\vr,t)) =
0, \label{eqn:macroeuler}
\end{equation}
where $\wz(\vr,t)$ and
$\u(\vr,t)$ 
are the macroscopic vorticity and the macroscopic velocity, respectively.
The point vortex system has been successfully applied
to the study of 2D turbulence \cite{Kraichnan, Tabeling}.
In the landmark paper published in 1949, Onsager proposed 
an application of statistical mechanics to the 2D point 
vortex system, in which he sketched a possible explanation for 
the formation of large-scale, long-lived, vortex structures in 
turbulent flows \cite{Onsager, Eyink}. Negative temperature
equilibrium states described by the Boltzmann distribution (leading to the
sinh-Poisson
equation when considering the two-species point vortex system) are found by
Joyce and Montgomery \cite{Joyce}.
Since then, a large research effort has been devoted to understand the
negative temperature states, both theoretically and numerically
\cite{Montgomery1974, Kida1975, Kraichnan1975, Seyler1976, Pointin1976,
Lundgren1977, Ting1987, Robert1991, Eyink1993, Buehler2002,
Yatsuyanagi2005,Chavanis2002}.
On the other hand, it has been pointed out that a decaying 2D 
Navier-Stokes turbulence reaches an equilibrium state described 
by the sinh-Poisson equation \cite{Matthaeus1991, Montgomery1993, Li1996}.

Here, a question arises.
A distribution of the point vortices $\hwz(\vr,0)$ 
is given at $t=0$.
A time-evolved distribution $\hwz(\vr, T)$ 
at a certain time $T$ 
is obtained by solving the microscopic Euler equation (\ref{eqn:euler}).
On the other hand, a macroscopic vorticity field $\omega(\vr,0)$ at $t=0$ 
is obtained by a space average of $\hwz(\vr, 0)$, 
namely $\omega(\vr,0) = \langle \hwz(\vr,0) \rangle$.
Of course, a time-evolved macroscopic vorticity field
$\omega(\vr, T)$ is obtained by solving the macroscopic Euler equation 
(\ref{eqn:macroeuler}).
Is the space-averaged point vortex solution $\langle \hwz(\vr,T) \rangle$
the same as the macroscopic vorticity field $\omega(\vr,T)$?
If the number of point vortices $N \rightarrow +\infty$, the
answer is ``yes'' because there is no fluctuation about the mean
field \cite{Note01}.
By
contrast, for finite $N$ systems, the answer
is ``no'' because there are fluctuations implying a deviation to
the macroscopic Euler equation. In that
case,  the
evolving equation
for $\omega(\vr,{t})$ which is exactly equal to $\langle
\hwz(\vr,{t}) \rangle$ 
should be written as 
\begin{equation}
	\frac{\partial \wz(\vr,t)}{\partial t} + \div (\u(\vr,t) \wz(\vr,t)) =
C, \label{eqn:macro-with-collision}
\end{equation}
where $C$ is a collision term. In the following, we restrict our discussion to
determining an
explicit formula of $C$, i.e., we develop a kinetic theory
of point vortices.

The kinetic theory of point vortices has attracted a lot of
attention. Let us briefly review earlier works on the subject. A general kinetic
equation for point vortices, valid for arbitrary flows
(axisymmetric or not), has been obtained by Chavanis
\cite{Chavanis2001,Chavanis2008} with several equivalent
methods (projection operator technics, the BBGKY hierarchy, and
the Klimontovich
approach). It writes \cite{Note02}
\begin{eqnarray}
	\frac{\partial \omega(\vec{r},t)}{\partial t}+\nabla\cdot( \vec{u}(
	\vec{r},t)\omega(\vec{r},t))
	=\frac{\partial}{\partial
	{r}^{\mu}}\int_0^{+\infty} d\tau \int d\vec{r}_{1}
	{V}^{\mu}(1\rightarrow
	0)G(t,t-\tau)\nonumber\\
	\times  \left \lbrack {{\tilde V}}^{\nu}(1\rightarrow 0) {\partial\over\partial
	{ r}^{\nu}}+{{\tilde V}}^{\nu}(0\rightarrow 1) {\partial\over\partial
	{r}_{1}^{\nu}}\right \rbrack 
	{\omega}(\vec{r},t)
	\frac{\omega}{\Omega}(\vec{r}_1,t), \label{gen7}
\end{eqnarray}
where $G(t,t-\tau)$ is a Green function constructed 
with the mean velocity, $\vec{V}(i\rightarrow j)$ is the
velocity created by point vortex $i$ on point vortex $j$,
$\tilde{\vec{V}}(i\rightarrow j)$ is the fluctuating velocity, and $\Omega$ is the
circulation of a point vortex. This equation is valid at the order $O(1/N)$
when $N\rightarrow +\infty$, so
it describes the evolution of the system of point vortices, due to two-body
distant encounters, on a time scale of the order $Nt_D$ where
$t_D$ is the dynamical
time.  For axisymmetric flows, the point vortices have
a circular motion with
angular velocity $\Omega(r,t)$. In that case, the collision term can be
simplified and the kinetic equation (\ref{gen7}) takes the form
\cite{Chavanis2001,Chavanis2008}:
\begin{eqnarray}
	\label{sd4}
	\frac{\partial\omega(r,t)}{\partial t}=-\frac{\Omega}{4}\frac{1}{r}\frac{\partial}{\partial r}\int_0^{+\infty}r'\,
	dr' \ln\left\lbrack 1-\left (\frac{r_{<}}{r_{>}}\right )^2\right\rbrack
	\delta(\Omega(r,t)-\Omega(r',t))\nonumber\\
	\times\left (\frac{1}{r}\frac{\partial}{\partial
	r}-\frac{1}{r'}\frac{\partial}{\partial r'}\right
)\omega(r,t)\omega(r',t).
\end{eqnarray}
This equation conserves circulation, energy, angular momentum, 
and it increases monotonically the Boltzmann entropy ($H$ theorem). The
collisional evolution of the point vortices is due to a condition of resonance
encapsulated in the $\delta$-function. We note that when the profile of angular
velocity is, or becomes, monotonic, the condition of resonance cannot be
satisfied anymore and the collision term vanishes. As a result, the evolution
described by the kinetic equation (\ref{sd4}) stops even if the attained
distribution differs from the Boltzmann distribution (this is because the
kinetic equation (\ref{sd4}) admits an infinity of steady states in addition to
the Boltzmann distribution,
namely any distribution with a monotonic profile of angular velocity).
Therefore, the kinetic equation (\ref{sd4}) usually does {\it not} converge
towards the
Boltzmann distribution \cite{cl}. This is not a problem. It simply tells us
that, for
axisymmetric flows, the relaxation towards the Boltzmann distribution is
governed by another kinetic equation, valid at the order $1/N^2$ (or at a
higher order), taking into account more complicated correlations between point
vortices than simply two-body collisions. As a result, the relaxation time
towards the Boltzmann distribution is of order $N^2 t_D$ or longer \cite{Chavanis2012,Chavanis2012b,cl,Chavanis2010}. It is also
possible that the point vortex gas (in the axisymmetric situation) never
achieves the Boltzmann distribution. This is still an open problem. For
non-axisymmetric flows, a natural strategy would be to introduce
angle-action variables to obtain a generalization of the kinetic equation
(\ref{sd4}) similarly to
what has been done in the context of the Landau and  Lenard-Balescu equations in
stellar
dynamics \cite{Chavanis2010,Heyvaerts,Chavanis2012c,Chavanis2013}. Alternatively,
Chavanis \cite{Chavanis2001,Chavanis2008}  has proposed a
heuristic simplification of the
kinetic equation (\ref{gen7}) in the form
\begin{eqnarray}
\frac{\partial {\omega}}{\partial t}+{\vec{u}}\cdot
\nabla {\omega}=\frac{\Omega}{8}
\nabla \cdot \int d\vec{r}'\, \frac{\vec{x}_{\perp}\otimes
\vec{x}_{\perp}}{x^2}\delta({\vec{x}}\cdot {\vec{w}})\left ( {\omega}'
\nabla{\omega}- {\omega} \nabla' {\omega}'\right ),
\label{grd}
\end{eqnarray}
where $\omega=\omega(\vec{r},t)$, $\omega'=\omega(\vec{r}',t)$, 
$\vec{x}=\vec{r}-\vec{r}'$, and $\vec{w}=\vec{u}(\vec{r},t)-\vec{u}(
\vec{r}',t)$. This equation conserves circulation, energy, angular momentum, 
and it increases monotonically the Boltzmann entropy ($H$ theorem). Furthermore,
for axisymmetric flows, it reduces to a form very close to the exact Eq.
(\ref{sd4}) up to logarithmic corrections. As in Eq. (\ref{sd4}), the
collisional evolution of the
point vortices according to Eq. (\ref{grd}) is due to a condition of resonance
encapsulated in the
$\delta$-function. However, this condition of resonance is more complicated
(hence more easily satisfied) than in Eq. (\ref{sd4}) so we may expect that the
solution
of this equation reaches, or approaches, the Boltzmann distribution on the
time scale $Nt_D$ over which this equation is valid. Indeed, there are numerical
observations that, for non-axisymmetric flows, the relaxation time is of order
$Nt_D$ \cite{kn}. Therefore, if there are enough resonances, the kinetic
equation (\ref{grd}) will drive the system towards the Boltzmann distribution
on a
time scale $Nt_D$ while if the resonances cannot be satisfied the kinetic
equation (\ref{grd}) is not sufficient to describe the dynamics (we have to
take terms
of order $1/N^2$, or higher, into account) and the relaxation time will be of
order $N^2t_D$, or longer \cite{Chavanis2012,Chavanis2012b,cl,Chavanis2010}. As discussed in Refs. \cite{Chavanis2001,Chavanis2008},  
the kinetic
equations (\ref{gen7})-(\ref{grd})
have the form of Fokker-Planck equations
\begin{eqnarray}
\frac{\partial {\omega}}{\partial t}+\vec{u}\cdot
\nabla {\omega} = \nabla \cdot ({\sf D}\cdot \nabla\omega -\omega \vec{V}),
\label{fp}
\end{eqnarray}
including a diffusion term and a drift term (they exactly reduce
to Fokker-Planck equations in the test particle approach). The drift term  was
first evidenced in Ref. \cite{chavanis1998} and it plays a fundamental role in the
kinetic theory of
point vortices in relation to the process of self-organization.
Finally, these kinetic equations can be easily generalized to the multi-species
point vortex
gas as discussed in Refs. \cite{Dubin2003,cl}.

In the previous kinetic theories 
\cite{Chavanis2001,Chavanis2008,DubinONeil1988,Dubin2003,Chavanis2010,Chavanis2012,Chavanis2012b,cl,chavanis1998}, 
the mean field is assumed to be ``strong'' 
and the fluctuations ``weak'' so that the point vortices are advected by the 
mean velocity for a long time. For example, in an axisymmetric flow, the point
vortices follow circular trajectories in a first approximation slightly
perturbed by the effect of the long-range collisions (whose strength is of order
$|\delta {\vec{u}}|\sim 1/\sqrt{N}$) on a very long time scale. 
In that case, the dynamics of the point vortices is ``resonant" (see the 
$\delta$-function in Eqs. (\ref{sd4}) and (\ref{grd})). 
The opposite situation is when there is no mean flow. 
This is the situation investigated by Taylor and McNamara \cite{TaylorMcNamara1971} 
and Dawson and collaborators \cite{DawsonOkudaCarlile1971,OkudaDawson1973}. 
They consider a neutral system of point vortices at equilibrium where the vortices 
are uniformly distributed in average.
In that case, there is no kinetic equation for $\omega(\vr,t)$ since $\omega(\vr,t)=0$ 
at each time $t$ (however, there can be large-scale fluctuations giving rise to Dawson 
vortices). 
Taylor and McNamara \cite{TaylorMcNamara1971} consider the diffusion of a test 
vortex in a uniform vorticity background
where the field vortices can be correlated (having a thermal distribution) 
or uncorrelated (having a random distribution) and derive the 
corresponding diffusion coefficient.
In that case, there is no drift since the vorticity is uniform. 
In the present paper, we consider a situation intermediate between the works of 
Refs. \cite{Chavanis2001,Chavanis2008,DubinONeil1988,Dubin2003,Chavanis2010,Chavanis2012,Chavanis2012b,cl,chavanis1998}
and the works of Refs. \cite{TaylorMcNamara1971,DawsonOkudaCarlile1971,OkudaDawson1973}.
We assume that  the mean field is weak, but non-zero,
so the time scale of the
macroscopic motion
is long as compared to the decorrelation time.
In this case, we can make a linear trajectory approximation with the local 
velocity field ${\vec{u}}({\vec{r}},t)$ to compute the correlation function.
In our problem, the distribution of the point vortices is spatially inhomogeneous 
and out of equilibrium, so our aim is to determine a kinetic equation that describes 
the relaxation of  $\omega(\vr,t)$ towards the Boltzmann distribution. 
Since the mean flow is weak, the kinetic evolution will not be ``resonant'' so it 
should be described by a kinetic equation different from Eqs. (\ref{sd4}) and (\ref{grd}).
This approach is not in contradiction with previous
works. It just explores a completely different regime, so it should be
considered as being complementary to previous works \cite{Note03}.
In Appendix \ref{sec_val}, we try to justify the domain of validity of
this approach by estimating the decorrelation time and the time scale of the
macroscopic motion.
It is shown that there is a critical value of $N_c$.
The case of $1 \ll N \ll N_c$ corresponds to the situation of the current paper
and the case of $N\gg N_c$ corresponds to the previous studies.
However, we make clear since the start
that the
present approach is not firmly justified mathematically (in a well-defined
asymptotic limit) and this is the reason of the problems encountered, and
discussed, at the end
of the paper. Despite these limitations, the linear trajectory approximation is
interesting in itself because it makes the kinetic theory very similar to that
developed in plasma physics and stellar dynamics where the particles have linear
trajectories due to their inertia. Therefore, it is interesting to see what a
similar approximation implies in the case of point vortices. Furthermore, it
leads to explicit kinetic equations that could be confronted to direct
numerical simulations. Actually, this linear trajectory approximation was
introduced in Appendix B of Ref. \cite{Chavanis2008} where the kinetic equation
(\ref{eqn:gamma}) was derived
directly from Eq. (\ref{gen7}). For self-consistency, this equation is
re-derived here from the start by using the Klimontovich formalism. Then, we
introduce a space average of the diffusion flux and derive the
kinetic
equation (\ref{finea}). The obtained collision term has the following good
properties.
(i) It conserves the mean field energy. (ii)
During the relaxation process towards the global equilibrium, the
system first reaches a local equilibrium state.
In the local equilibrium state, the relation 
$\omega_{\rm leq} = \omega_{\rm leq}[\psi]$
is satisfied in each small region inside which the inverse
temperature $\beta$ is
constant (notation $\omega[\psi]$ is a functional
of the stream function $\psi$).
In the small regions, the second term
$\div (\u \omega)$ of the Euler equation  (\ref{eqn:macro-with-collision}) 
vanishes.
Then the time evolution of the system is dominated by the collision
term.
However, the magnitude of the collision term is small compared
to that of $\div (\u\omega)$, and the speed of the relaxation
slows down.
When the system reaches a global equilibrium state described by
the Boltzmann distribution $\omega_{\rm eq} = \omega_0 
\exp(-\beta \Omega \psi_{\rm eq})$, the collision term completely
converges to zero all over the system and the Einstein relation is
obtained \cite{Chavanis2001}. (iii)
The obtained collision term satisfies the $H$ theorem
which guarantees that the system relaxes to a global equilibrium state. 
As the order of the collision
term is $O(1/N)$, the relaxation time scales as $N$.

In order to illustrate our theoretical study, we have performed 
direct  numerical simulations of point vortices  to confirm the scaling of the
relaxation time.
The results elucidate a new scaling according to which the relaxation time
$t_R$ is 
proportional to $1/\Omega$.
The previously obtained scaling $t_R \propto N t_D$ by Kawahara and 
Nakanishi \cite{kn} where $t_D$ is the dynamical time of the system is also
reexamined 
and confirmed.

The organization of this paper is as follows.
In Sec. II, the point vortex system and the Klimontovich formalism are briefly introduced.
In Sec. III, we demonstrate explicit formulae for the diffusion and the drift
terms as intermediate results.
In Sec. IV, a detailed calculation of the diffusion term is shown.
Since similar calculations can be applied to the drift term, the 
details for the drift term are omitted. In Sec. V, three good
properties of the collision term are demonstrated. 
In Sec. VI, we mention the limitation of our approach. Finally, in Sec. VII, we
compare our
results with direct numerical simulations of point vortices. We
find a good qualitative agreement.
\section{Point Vortex System}
Consider a 2D system consisting of $N$ positive point vortices \cite{Newton}.
The circulation of each point vortex is given by a positive constant $\Omega$. Therefore,
\begin{equation}
	\hwz = \sum_i^{N} \Omega \delta(\vr - \vr_i), \label{eqn:point-vortex}
\end{equation}
where $\hwz = \hwz(\vr,t)$ is the $z$-component of the microscopic vorticity on the $x-y$ plane,
and $\delta(\vr)$ is the Dirac delta function in two dimensions.
The microscopic variables in the microscopic equation are
identified by $\hat{\cdot}$.
For brevity, we shall omit the $t$ and $\vr$ dependences if there is no
ambiguity.
Vector $\vr_i = \vr_i(t)$ is the position vector of the $i$-th point vortex.
The discretized vorticity (\ref{eqn:point-vortex}) is a formal solution
of the microscopic Euler equation (\ref{eqn:euler}).
Other microscopic variables are defined by
\begin{eqnarray}
	&&\hat{\u}  =  \hat{\u}(\vr,t) = -\hat{\vec{z}} \times \grad \hat{\psi},
\\
	&&\hat{\psi}  =  \hat{\psi}(\vr,t) = \sum_i\Omega_iG(\vr - \vr_i), \\
	&&G(\vr)  =  -\frac{1}{2\pi} \ln|\vr|,
\end{eqnarray}
where $\hat{\u}$ and $\hat{\psi}$ are the velocity field and the stream 
function in the 2D plane, 
$\hat{\vec{z}}$ is the unit vector in the $z$-direction, 
and $G(\vr)$ is the 2D Green function for the Laplacian
operator in an infinite domain. Since the solution of a macroscopic fluid
equation should be given by a smooth function,
the singular solution (\ref{eqn:point-vortex}) should be regarded not as
a solution of the macroscopic equation but as one of the microscopic equation.
Thus we call the equation that has the microscopic point vortex solution (\ref{eqn:point-vortex}), 
the ``microscopic" Euler equation.

There exist many analogies between point vortices, plasmas, and 
stellar systems (despite, of course, some important differences), and
similar methods can be developed to study these systems \cite{Chavanis2002}.
In plasma physics and stellar dynamics, the
evolution of the macroscopic phase space density
$f({\vr},{\vec{v}},t)$ is
governed by the
Vlasov-Landau kinetic equation
\begin{eqnarray}
\frac{\partial f}{\partial t}+\vec{v}\cdot \nabla_{\vec{r}} f+\vec{F}\cdot
\nabla_{\vec{v}} f=A\frac{\partial}{\partial v_i}\int d\vec{v}'
\frac{w^2\delta_{ij}-w_iw_j}{w^3}\left (f'\frac{\partial f}{\partial
v_j}-f\frac{\partial f'}{\partial v'_j}\right ),
\label{exp5}
\end{eqnarray}
where $f=f(\vec{r},\vec{v},t)$, $f'=f(\vec{r},\vec{v}',t)$, 
$\vec{w}=\vec{v}-\vec{v}'$, $\vec{F}(\vec{r},t)$
is the mean force by unit of mass acting on a particle, and $A$ is a constant
($A=2\pi m
G^2\ln\Lambda$ for stellar systems and $A=(2\pi e^4/m^3)\ln\Lambda$ for plasmas
where $\ln\Lambda$ is the Coulombian logarithm). This equation can be
derived from the Klimontovich equation for the microscopic phase space density
$\hat{f}$ that is
\begin{equation}
	\frac{\partial \hat{f}}{\partial t} + \vec{v}\cdot
\grad_{\vec{r}}\hat{f} +
\hat{\vec{F}}\cdot \grad_{\vec{v}}\hat{f} = 0
\label{klim}
\end{equation}
by using a quasilinear approximation \cite{Klimontovich}. The
Vlasov-Landau equation has the form of a Fokker-Planck equation
\begin{eqnarray}
\frac{\partial f}{\partial t}+\vec{v}\cdot \nabla_{\vec{r}} f+\vec{F}\cdot
\nabla_{\vec{v}} f=\nabla_{\vec{v}}\cdot ({\sf D}\cdot \nabla_{\vec{v}} f-f \vec{A}),
\label{fpbi}
\end{eqnarray}
including a diffusion term and a friction (it exactly reduces 
to a Fokker-Planck equation in the test particle approach). As the dynamics of
plasmas and stellar systems is usually dominated not by collisions but
by a collective behavior due
to long-range interactions, the collision term can often be neglected
(on a
collisionless  time scale or for $N\rightarrow +\infty$) and it yields the
simplest
form of the kinetic equation called the Vlasov equation:
\begin{equation}
	\frac{\partial f}{\partial t} + \vec{v}\cdot \grad_{\vec{r}}f +
\vec{F}\cdot \grad_{\vec{v}}f = 0.
\label{vlasov}
\end{equation}

The same hierarchy exists in the 2D fluid equations \cite{Chavanis2002}.
The most microscopic equation is the microscopic Euler equation
(\ref{eqn:euler}),
which has the discrete particle solution
(\ref{eqn:point-vortex}). This is the counterpart of the
Klimontovich equation (\ref{klim})  in plasma physics and stellar dynamics.
Dividing the microscopic variables into a macroscopic and a fluctuation part, and 
taking ensemble average yields a macroscopic fluid equation with a collisional
effect like the kinetic equations (\ref{gen7})-(\ref{grd})
derived in the past
or like the kinetic equation (\ref{finea}) derived in this paper.
These kinetic equations
are analogous to the Vlasov-Landau equation (\ref{exp5}). They have a Fokker-Planck
form in which the drift of the point vortices  is
the counterpart of the dynamical friction [compare Eqs. (\ref{fp}) and (\ref{fpbi})], as noted in
Ref. \cite{chavanis1998}. Ignoring the collision term in the above macroscopic
equations, we obtain 
the inviscid fluid equation, namely the macroscopic Euler equation (\ref{eqn:macroeuler}).
This is the counterpart of the Vlasov equation (\ref{vlasov}). These analogies
were first pointed out in Ref. \cite{Chavanis2002}.

The starting equation is the microscopic Euler equation (\ref{eqn:euler}).
Inserting the following expressions into Eq. (\ref{eqn:euler}),
\begin{eqnarray}
	\hwz          & = & \wz     + \delta \wz, \label{eqn:micro-w}\\
	\hat{\vec{u}} & = & \vec{u} + \delta \vec{u},\label{eqn:micro-u}
\end{eqnarray}
and taking the ensemble average, we obtain the
following macroscopic 
equation with the collision term $C=C(\vr, t)$
\begin{equation}
	\frac{\partial \wz}{\partial t} + \div(\u \wz) = C,\label{eqn:order1}
\label{eqn:collision-term-org}
\end{equation}
where
\begin{eqnarray}
	& &C  \equiv  -\div \vec{\Gamma}(\vr, t), \\
	& &\vec{\Gamma}  =  \langle \delta\u\delta \wz\rangle 
          =  -\int d\vr' \vec{F}(\vr - \vr') \langle \dwz' \dwz\rangle,
\label{eqn:diffusion1}\\
	&&\vec{F}(\vr)  =  \hat{\vec{z}} \times \grad G(\vr),
\end{eqnarray}
where $\vec{\Gamma}$ denotes a diffusion flux.
We note $\dwz'$ for $\dwz(\vr',t)$.
Similarly, we shall note $\wz'$ for $\wz(\vr',t)$.
To obtain Eq. (\ref{eqn:diffusion1}), the  following relation has been utilized
\begin{equation}
	\delta \vec{u} =-\int d\vr' \vec{F}(\vr-\vr')\dwz'. \label{eqn:delta-u}
\end{equation}
In the next section, we will analytically assess the collision term $C$
in the case where a linear trajectory approximation can be
implemented.
\section{Evaluation of Collision Term}
We consider a point vortex system with large $N$ keeping the
total circulation $N\Omega$ constant. Therefore, $\Omega\sim 1/N$.
We expect that the collision term $C$ appearing in Eq. (\ref{eqn:collision-term-org})
for the point vortex system has two terms, a diffusion term proportional to $\grad \omega$ and 
a drift term proportional to $\omega$,
namely
\begin{equation}
	\vec{\Gamma} \equiv -{\sf D} \cdot \grad \omega + \vec{V} \omega,
\label{eqn:diffusion_flux}
\end{equation}
where ${\sf D} = {\sf D}(\vr,t)$ is a diffusion tensor 
and $\vec{V} = \vec{V}(\vr,t)$ is a drift velocity.
To evaluate $\sf D$ and $\vec{V}$ explicitly, we assume there exist a small
parameter $\epsilon$ such that:
\begin{eqnarray}
	& & \omega \approx \grad^2\psi \approx O(\epsilon^0),\quad \vec{u} \approx \grad \psi \approx O(\epsilon^0),\quad \grad\vec{u} \approx \grad^2\psi \approx O(\epsilon^0),\nonumber\\
	& & \grad \omega \approx O(\epsilon^{1/2}),\quad \delta \omega \approx O(\epsilon^{1/2}),\quad \delta\vec{u} \approx O(\epsilon^{1/2}),\nonumber\\
	& & \frac{\partial \vec{u}}{\partial t} \approx O(\epsilon^{1/2}), \quad \frac{\partial \omega}{\partial t} \approx O(\epsilon^{1/2}), \quad \grad \grad \vec{u} \approx O(\epsilon^{1/2}),\nonumber\\
	& & \vec{\Gamma} \approx O(\epsilon).
\end{eqnarray}
The expansion parameter $\epsilon$ is similar to the one
introduced by Chavanis 
in Refs. \cite{Chavanis2001,Chavanis2008,Chavanis2012} and the references
therein. However, we assume here that the gradient of the vorticity
profile is weak. This is necessary for the validity of the linear trajectory
approximation. With this scaling, the left hand side of Eq. (\ref{eqn:order1})
is $O(\epsilon^{1/2})$, 
while the right hand side is $O(\epsilon^{3/2})$.
Expressing $\sf D$ and $\vec{V}$ in the form of a perturbation expansion and gathering
the terms of the appropriate order, an analytical
formula for the collision term $C$ will be obtained.

To rewrite the collision term in Eq. (\ref{eqn:collision-term-org}) according to
the above prospect, 
we introduce a linearized equation obtained by inserting 
Eqs. (\ref{eqn:micro-w}) and (\ref{eqn:micro-u})
into Eq. (\ref{eqn:euler}) and assembling the first-order fluctuation terms:
\begin{equation}
	\frac{\partial }{\partial t} \dwz + \div(\u\dwz) = -\delta\u \cdot \grad \wz. \label{eqn:linearlized}
\end{equation}
This is the counterpart of the quasilinear approximation in plasma physics \cite{Klimontovich}. As the macroscopic quantities $\vec{u}$ appearing in the second term in the 
left-hand side and $\grad \wz$ in the right-hand side are supposed to be 
constant in the time scale of the microscopic fluctuation,
Eq. (\ref{eqn:linearlized}) can be integrated:
\begin{eqnarray}
	\dwz & = & -\int_{t_0}^t d\tau \delta\u(\vr - \u(t-\tau),\tau) \cdot \grad\wz \nonumber\\
		& & + \dwz(\vr - \u(t - t_0),t_0), \label{eqn:integrated-linear-eq1}\\
	\dwz' & = & -\int_{t_0}^t d\tau \delta\u(\vr' - \u'(t-\tau),\tau) \cdot \grad'\wz' \nonumber\\
		& & + \dwz(\vr' - \u'(t - t_0),t_0), \label{eqn:integrated-linear-eq2}
\end{eqnarray}
where $\grad'\wz' = \grad_{\vr'}\wz(\vr')$.
This is called the linear trajectory approximation
where the trajectory of the point vortex is straight.
The validity of this approximation is discussed in Appendix
\ref{sec_val}.
The value of $t_0$ is chosen to satisfy $t - t_0 \gg t_c$ where $t_c$ is a 
correlation time of the fluctuations.
Substituting Eqs. (\ref{eqn:integrated-linear-eq1}) and (\ref{eqn:integrated-linear-eq2})
into the correlation term in Eq. (\ref{eqn:diffusion1}), we obtain
\begin{eqnarray}
	& & \left\langle \dwz' \dwz\right\rangle \nonumber\\
	& = & \left\langle \left( -\int_{t_0}^t d\tau \delta\u(\vr' - \u'(t-\tau),\tau) \cdot \grad' \wz' \right) \right.\nonumber\\
	& & 	\times \left.\left( -\int_{t_0}^t d\tau \delta\u(\vr - \u(t-\tau),\tau) \cdot \grad \wz \right) \right\rangle \nonumber\\
	&  & + \left\langle \left( -\int_{t_0}^t d\tau \delta\u(\vr' - \u'(t-\tau),\tau) \cdot \grad' \wz' \right) 
		\dwz(\vr - \u(t-t_0),t_0) \right\rangle \nonumber\\
	&  & + \left\langle \dwz(\vr' - \u'(t-t_0),t_0) 
		\left( -\int_{t_0}^t d\tau \delta\u(\vr - \u(t-\tau),\tau) \cdot \grad \wz \right)\right\rangle \nonumber\\
	&  & + \left\langle \dwz(\vr' - \u'(t-t_0),t_0) \dwz(\vr - \u(t-t_0),t_0) \right\rangle \label{eqn:diffusion2}\\
	& {\approx} & -\int_{t_0}^t d\tau \left\langle \delta
\u(\vr - \u(t-\tau) ,\tau) \dwz' \right\rangle \cdot \grad \wz \nonumber\\
	& & - \int_{t_0}^t d\tau \left\langle \delta \u(\vr' - \u'(t-\tau)
,\tau) \dwz \right\rangle \cdot \grad' \wz' \label{eqn:diffusion3}\\
	& \approx & \int_{t_0}^t d\tau \int d\vr'' \vec{F}(\vr - \u(t-\tau)- \vr'') \cdot \grad \wz \left\langle \dwz(\vr'',\tau) \dwz'\right\rangle  \nonumber\\
	& &  +\int_{t_0}^t d\tau \int d\vr'' \vec{F}(\vr' - \u'(t-\tau)-\vr'') \cdot \grad' \wz' \left\langle \dwz(\vr'',\tau) \dwz\right\rangle. \label{eqn:diffusion4}
\end{eqnarray}
When obtaining formula (\ref{eqn:diffusion3}), we assume that the first 
term in formula (\ref{eqn:diffusion2}) is negligible as it has two nablas.
We drop the last term as it should have a factor of $1/(t-t_0)$ and we focus on
the $t-t_0\gg t_c$ case.
The time is shifted from $t_0$ to $t$ using the linear
trajectory approximation.
When rewriting formula (\ref{eqn:diffusion3}) as (\ref{eqn:diffusion4}), 
Eq. (\ref{eqn:delta-u}) is used.

Inserting Eq. (\ref{eqn:diffusion4}) into Eq. (\ref{eqn:diffusion1}), the following intermediate results
are obtained:
\begin{eqnarray}
	& & {\sf D} \cdot \grad \wz \nonumber\\
	& = & \int_{t_0}^t d\tau\int d\vr'\int d\vr''\vec{F}(\vr-\vr') \vec{F}(\vr - \u(t-\tau) - \vr'') \cdot \grad \wz \nonumber\\
	& & 	\times \left\langle \dwz(\vr'',\tau) \dwz'\right\rangle, \label{eqn:diffusion}\\
	& & \vec{V}\wz\nonumber\\
	& = & -\int_{t_0}^t d\tau\int d\vr'\int d\vr''\vec{F}(\vr-\vr') \vec{F}(\vr' - \u'(t-\tau) - \vr'') \cdot \grad' \wz' \nonumber\\
	& & 	\times \left\langle \dwz(\vr'',\tau) \dwz\right\rangle. \label{eqn:friction}
\end{eqnarray}
It should be noted that the diffusion term can be expressed in the following
modified Kubo formula:
\begin{equation}
	{\sf D} = \int_{t_0}^t d\tau \langle \delta\u(\vr,t)\delta\u(\vr-\u(t-\tau),\tau) \rangle.
\end{equation}
\section{Evaluation of Diffusion and Drift Terms}
As the expression of the diffusion term (\ref{eqn:diffusion}) is very similar to that of the drift term
(\ref{eqn:friction}), the detailed derivation for diffusion term only is shown.
We start with
\begin{eqnarray}
	& & \left\langle \dwz(\vr'',\tau) \dwz'\right\rangle\nonumber\\
	& = & \left\langle \left[ \hwz(\vr'',\tau) -\wz(\vr'',\tau) \right] \left[ \hwz' - \wz' \right] \right\rangle\nonumber\\
	& = & \left\langle \hwz(\vr'',\tau) \hwz'\right \rangle - \wz(\vr'',\tau) \wz' \nonumber\\
	& = & \left\langle \sum_{i=1}^{N} \Omega^2 \delta(\vr''-\vr_i(\tau))\delta(\vr'-\vr_i(t))\right\rangle\nonumber\\
	& & + \left\langle \sum_{i=1}^{N} \sum_{j\neq i}^{N}\Omega^2 \delta(\vr''-\vr_i(\tau)) \delta(\vr'-\vr_j(t))\right\rangle\nonumber\\
	& & - \wz(\vr'',\tau) \wz'.\label{eqn:diffusion6}
\end{eqnarray}
The first term in the last result in Eq. (\ref{eqn:diffusion6}) corresponds to
the case of $i=j$,
and the second term corresponds to the case of $i\neq j$.

For the $i=j$ case, the formula is rewritten as
\begin{eqnarray}
	& & \left\langle \sum_{i=1}^{N} \Omega^2 \delta(\vr''-\vr_i(\tau))\delta(\vr'-\vr_i(t))\right\rangle \nonumber\\
	& = & \left\langle \sum_{i=1}^{N} \Omega^2 \delta(\vr'' - \vr_i(\tau) - \vr' + \vr_i(t)) \delta(\vr'-\vr_i(t))\right\rangle\nonumber\\
	& = & \sum_{i=1}^{N} \Omega^2 \left\langle \delta(\vr'' - \vr_i(\tau) - \vr' + \vr_i(t)) \delta(\vr'-\vr_i(t))\right\rangle. \label{eqn:diffusion7}
\end{eqnarray}
Here we introduce a stochastic process to evaluate $\vr_i(t) - \vr_i(\tau)$:
\begin{eqnarray}
	\vr_i(t)-\vr_i(\tau) & = &
\int_{\tau}^t\u(\vr_{i}(\tau'),\tau')d\tau' + \vec{\xi}_i
\nonumber\\
	& \approx & \u_{i}(t-\tau) + \vec{\xi}_i.
\label{eqn:diffusion8}
\end{eqnarray}
The first term in Eq. (\ref{eqn:diffusion8}) represents the
linear trajectory
approximation and the second term represents a Brownian motion.
The stochastic process represented by $\langle \cdot \rangle_{\xi}$ includes all the possible motion
to reach position $\vr_i$ at time $t$.
Then, Eq. (\ref{eqn:diffusion7}) can be rewritten as
\begin{eqnarray}
	& & \sum_{i=1}^{N} \Omega^2 \left\langle \delta(\vr'' - \vr_i(\tau) - \vr' + \vr_i(t)) \delta(\vr'-\vr_i(t))\right\rangle \nonumber\\
	& = & \sum_{i=1}^{N} \Omega^2 \left\langle
\delta(\vr''-\vr'+{\u_i}(t-\tau) +
\vec{\xi}_i)\right\rangle_{\xi} \left\langle \delta(\vr' -
\vr_i(t))\right\rangle\nonumber\\
	& = & \left\langle \delta(\vr''-\vr' + \u'(t-\tau)+
\vec{\xi})\right\rangle_{\xi}\Omega \wz'.
\end{eqnarray}

For the $i\neq j$ case, we introduce an approximation valid for large $N$ that correlations between 
point vortices
can be neglected
\begin{eqnarray}
	& & \sum_{i}^{N} \sum_{j \neq i}^{N} \Omega^2 \left\langle \delta(\vr'' - \vr_i(\tau)) \delta(\vr' - \vr_j(t))\right\rangle\nonumber\\
	& \approx & \sum_i^{N} \sum_{j \neq i}^{N} \Omega^2 \left\langle \delta(\vr''-\vr_i(\tau))\right\rangle \left\langle \delta(\vr' - \vr_j(t)) \right\rangle. \label{eqn:diffusion9}
\end{eqnarray}
We also use the following relation:
\begin{equation}
	\sum_i^{N}\Omega \left\langle \delta(\vr''-\vr_i(\tau))\right\rangle =
\wz(\vr'',\tau). \label{eqn:approx1}\label{eqn:approx2}
\end{equation}
Inserting Eq. (\ref{eqn:approx2}) 
into Eq. (\ref{eqn:diffusion9}), we obtain
\begin{eqnarray}
	& & \sum_i^{N} \sum_{j \neq i}^{N} \Omega^2 \left\langle \delta(\vr''-\vr_i(\tau))\right\rangle \left\langle \delta(\vr' - \vr_j(t)) \right\rangle \nonumber\\
	& = & \left( \sum_{i=1}^{N} \Omega \left\langle \delta(\vr''-\vr_i(\tau))\right\rangle\right) 
		\times \left( \sum_{j \neq i}^{N}\Omega \left\langle \delta(\vr'-\vr_j(t))\right\rangle\right) \nonumber\\
	& = & N\Omega \frac{\wz(\vr'',\tau)}{N\Omega}(N-1)\Omega\frac{\wz'}{N\Omega}\nonumber\\
	& = & \wz(\vr'',\tau) \wz' - \frac{1}{N}\wz(\vr'',\tau)\wz'.
\end{eqnarray}
Combining the results of $i=j$ and $i\neq j$ cases, we rewrite Eq. (\ref{eqn:diffusion6}) as
\begin{eqnarray}
	& & \left\langle \dwz(\vr'',\tau) \dwz(\vr',t)\right\rangle\nonumber\\
	& = & \Omega \left\langle \delta(\vr'' - \vr' + \u'(t-\tau) +
\vec{\xi}) \right\rangle_{\xi} \wz' \nonumber\\
	& & - \frac{1}{N} \wz(\vr'',\tau)\wz'. \label{eqn:diffusion10}
\end{eqnarray}

The two terms in the right hand side of Eq. (\ref{eqn:diffusion10}) are of the
same order since
we request the total circulation $N\Omega$ to be constant.
To proceed with the evaluation of these terms, a conservation law is introduced
\begin{equation}
	\int d\vr' \left\langle \dwz(\vr'',\tau) \dwz'\right\rangle = 0. \label{eqn:conservation-law}
\end{equation}
Inserting Eq. (\ref{eqn:diffusion10}) into Eq. (\ref{eqn:conservation-law}), we obtain
\begin{eqnarray}
	& & \int d\vr' \left[ \Omega \left\langle \delta(\vr'' - \vr' +
\u'(t-\tau) + \vec{\xi}) \right\rangle_{\xi} \wz'\right. \nonumber\\
	& & -\left.\frac{1}{N}\wz(\vr'',\tau) \wz'\right] \nonumber\\
	& = & \Omega \int d\vr' \left\langle \delta(\vr''-\vr'+\u'(t-\tau) +
\vec{\xi})\right\rangle_{\xi}\wz' \nonumber\\
	& & -\frac{1}{N}\wz(\vr'',\tau)\int d\vr' \wz' \nonumber\\
	& = & \Omega \int d\vr' \left\langle \delta(\vr''-\vr'+\u'(t-\tau) +
\vec{\xi})\right\rangle_{\xi} \wz' \nonumber\\
	& & - \frac{1}{N}\wz(\vr'',\tau)N\Omega \nonumber\\
	& = & 0.
\end{eqnarray}
This equation yields 
\begin{equation}
	\wz(\vr'',\tau) = \int d\vq'\left\langle
\delta(\vr''-\vq'+\u(\vq')(t-\tau) + \vec{\xi})\right\rangle_{\xi}\wz(\vq',t),
\label{eqn:wr''}\\
\end{equation}
where $d\vr'$ is replaced by $d\vq'$ to avoid ambiguity.
This equation enables that all the quantities at $\tau$ are converted by ones at
$t$.
Inserting Eqs. (\ref{eqn:diffusion10}) and (\ref{eqn:wr''}) into Eq. (\ref{eqn:diffusion}),
we obtain
\begin{eqnarray}
 	& & {\sf D} \cdot \grad \wz \nonumber\\
	& = & \Omega \int_{t_0}^t d\tau\int d\vr'\int d\vr''\vec{F}(\vr-\vr') \vec{F}(\vr - \u(t-\tau) - \vr'') \cdot \grad \wz  \nonumber\\
	& & \times \left\langle \delta(\vr''-\vr' + \u'(t-\tau)+
\vec{\xi})\right\rangle_{\xi} \wz'\nonumber\\
	& & - \frac{1}{N} \int_{t_0}^t d\tau \int d\vr' \int d\vr'' \vec{F}(\vr - \vr') \vec{F}(\vr - \u(t-\tau) - \vr'')\cdot \grad \wz\nonumber\\
	& & \times \wz' \int d\vq' \left\langle
\delta(\vr''-\vq'+\u(\vq')(t-\tau) + \vec{\xi})\right\rangle_{\xi}\wz(\vq',t).
\label{eqn:diffusion11}
\end{eqnarray}
We proceed with the evaluation of the second term in Eq. (\ref{eqn:diffusion11}):
\begin{eqnarray}
	& & - \frac{1}{N} \int_{t_0}^t d\tau \int d\vr' \int d\vr'' \vec{F}(\vr - \vr') \vec{F}(\vr - \u(t-\tau) - \vr'')\cdot \grad \wz\nonumber\\
	& & \times \wz' \int d\vq' \left\langle
\delta(\vr''-\vq'+\u(\vq')(t-\tau) + \vec{\xi})\right\rangle_{\xi}\wz(\vq',t)
\nonumber\\
	& = & - \frac{1}{N} \int d\vr' \vec{F}(\vr - \vr') \int_{t_0}^t d\tau
\int d\vq'\left\langle \vec{F}(\vr - \vq' - (\u - \u(\vq'))(t-\tau) +
\vec{\xi})\right\rangle_{\xi} \cdot \grad \wz\nonumber\\
	& & \qquad \times \wz' \wz(\vq',t) \label{eqn:diffusion12}\\
	& = & - \frac{1}{N} \int d\vr' \vec{F}(\vr - \vr') \int_{t_0}^t d\tau \int \frac{d\vk}{(2\pi)^2} \frac{\hat{\vec{z}} \times i\vk}{|\vk|^2} \cdot \grad \wz \nonumber\\
	& & \times\int d\vq' \exp\left[i\vk \cdot (\vr - \vq' - (\u -
\u(\vq'))(t-\tau))\right] \left\langle
\exp(i\vk\cdot\vec{\xi})\right\rangle_{\xi}\nonumber\\
	& & \quad \times \wz' \wz(\vq', t). \label{eqn:diffusion13}
\end{eqnarray}
To rewrite formula (\ref{eqn:diffusion12}) as (\ref{eqn:diffusion13}), we
have used the Fourier transformation:
\begin{eqnarray}
	& & \vec{F}(\vr - \vq' - (\u - \u(\vq'))(t-\tau)) \nonumber\\
	& = & \frac{1}{(2\pi)^2} \int d\vk\frac{\hat{\vec{z}}\times i\vk}{|\vk|^2}\exp(i\vk\cdot(\vr - \vq' - (\u - \u(\vq'))(t-\tau))).\label{eqn:fourier}
\end{eqnarray}
The term $\left\langle \exp(i\vk\cdot\vec{\xi})\right\rangle_{\xi}$ represents
a Brownian motion
of the point vortices with diffusion tensor ${\sf D}$ and is evaluated by the cumulant expansion:
\begin{eqnarray}
	\left\langle \exp(i\vk\cdot\vec{\xi})\right\rangle_{\xi} 
		& = & \exp\left(-  \frac{\vk \cdot {\sf D} \cdot \vk}{2}(t-\tau)\right) \nonumber\\
	& \equiv & \exp(-\nu(t-\tau)), \label{eqn:cumulant}
\end{eqnarray}
where $\nu$ is a small positive parameter.
Inserting the following formula into Eq. (\ref{eqn:diffusion13})
\begin{eqnarray}
	& & \int_{t_0}^t d\tau \exp\{[-i\vk\cdot(\u - \u(\vq'))- \nu](t-\tau)\}\nonumber\\
	& \approx & \pi\delta(\vk \cdot (\u - \u(\vq'))) - \frac{i\vk\cdot (\u - \u(\vq'))}{|\vk\cdot(\u-\u(\vq'))|^2 + \nu^2},\label{eqn:defofnu}
\end{eqnarray}
we obtain
\begin{eqnarray}
	&  & - \frac{1}{N} \int d\vr' \vec{F}(\vr - \vr') \int_{t_0}^t d\tau \int \frac{d\vk}{(2\pi)^2} \frac{\hat{\vec{z}} \times i\vk}{|\vk|^2} \cdot \grad \wz \nonumber\\
	& & \times\int d\vq' \exp\left[i\vk \cdot (\vr - \vq' - (\u -
\u(\vq'))(t-\tau))\right] \left\langle
\exp(i\vk\cdot\vec{\xi})\right\rangle_{\xi}\nonumber\\
	& & \quad \times \wz'\wz(\vq', t) \nonumber\\
	& = & - \frac{1}{N} \int d\vr' \vec{F}(\vr - \vr') \int \frac{d\vk}{(2\pi)^2} \frac{\hat{\vec{z}} \times i\vk}{|\vk|^2}\cdot \grad \wz \nonumber\\
	&  & \times\int d\vq' \left[ \pi\delta(\vk\cdot(\u-\u(\vq')) - \frac{i\vk\cdot(\u-\u(\vq'))}{|\vk\cdot(\u-\u(\vq'))|^2 + \nu^2}\right]\exp(i\vk\cdot(\vr - \vq'))\nonumber\\
	& & \quad \times \wz' \wz(\vq',t).\label{eqn:diffusion14}
\end{eqnarray}

We substitute $\vr+\vq''$ for $\vq'$ and expand $\vec{u}(\vq')$ and $\wz(\vq')$ 
in the form of Taylor series and retain the zero-th order terms only:
\begin{eqnarray}
	\u(\vq') & {\approx} & \u(\vr) + \vq''\cdot \grad \u(\vr),
\label{eqn:taylor1}\\
	\wz(\vq') & = & \wz(\vr+\vq'') \approx \wz(\vr).\label{eqn:taylor2}
\end{eqnarray}
Inserting Eqs. (\ref{eqn:taylor1}) and (\ref{eqn:taylor2}) into Eq. (\ref{eqn:diffusion14}),
we finally obtain
\begin{eqnarray}
	& & - \frac{1}{N} \int d\vr' \vec{F}(\vr - \vr') \int \frac{d\vk}{(2\pi)^2} \frac{\hat{\vec{z}} \times i\vk}{|\vk|^2}\cdot \grad \wz \nonumber\\
	&  & \times \int d\vq' \left[ \pi\delta(\vk\cdot(\u-\u(\vq')) - \frac{i\vk\cdot(\u-\u(\vq'))}{|\vk\cdot(\u-\u(\vq'))|^2 + \nu^2}\right]\exp(i\vk\cdot(\vr - \vq'))\nonumber\\
	& & \quad \times \wz' \wz(\vq',t) \nonumber\\
	& = & - \frac{1}{N} \int d\vr' \vec{F}(\vr - \vr') \int \frac{d\vk}{(2\pi)^2} \frac{\hat{\vec{z}} \times i\vk}{|\vk|^2}\cdot \grad \wz\nonumber\\
	& & \times \int d\vq''\left[ \pi\delta(-\vk \cdot (\vq''\cdot\grad) \u) - \frac{i\vk\cdot(\vq''\cdot \grad)\u}{|\vk \cdot (\vq'' \cdot \grad)\u|^2 + \nu^2}\right]\exp(-i\vk \cdot \vq'')\nonumber\\
	& & \quad \times \wz' \wz. \label{eqn:diffusion15}
\end{eqnarray}
It is found that Eq. (\ref{eqn:diffusion15}) changes its sign under the transformation
$\vec{k} \rightarrow  -\vec{k}$ and $\vq'' \rightarrow  -\vq''$. Thus it is concluded that the 
integral equals zero, i.e. the second term in Eq. (\ref{eqn:diffusion11})
has zero contribution and only the first term remains. Repeating the above
procedure,
the obtained formula for the diffusion term is as follows:
\begin{eqnarray}
	& & {\sf D}\cdot \grad \wz\nonumber\\
	& = & \Omega\int d\vr'\vec{F}(\vr - \vr')\int \frac{d\vk}{(2\pi)^2}\exp(i\vk\cdot(\vr - \vr')) \frac{\hat{\vec{z}}\times i\vk}{|\vk|^2}\cdot \wz' \grad \wz\nonumber\\
	& & \times\left[\pi\delta(\vk \cdot (\u - \u')) - \frac{i\vk\cdot(\u-\u')}{|\vk \cdot (\u - \u')|^2 + \nu^2}\right].
\end{eqnarray}
A similar calculation can be adapted for the drift term.
For this case, the following conservation law is used:
\begin{equation}
	 \int d\vr \left\langle\dwz(\vr'',\tau) \dwz(\vr,t) \right\rangle=0.
\end{equation}

The whole result including both diffusion and drift terms is given by
\begin{eqnarray}
	\vec{\Gamma}& =& -{\sf D}\cdot \grad \wz + \vec{V}\wz \nonumber\\
	& = & -\Omega \int d\vr'\int \frac{d\vk}{(2\pi)^2} \int \frac{d\vk'}{(2\pi)^2}
		\exp(i(\vk+\vk')\cdot(\vr - \vr'))\nonumber\\
	& & \times \left[\pi\delta(\vk \cdot (\u - \u')) - \frac{i\vk\cdot(\u-\u')}{|\vk \cdot (\u - \u')|^2 + \nu^2}\right]\nonumber\\
	& & \times \frac{\hat{\vec{z}} \times i\vk'}{|\vk'|^2}
\frac{\hat{\vec{z}} \times i\vk}{|\vk|^2} \cdot \left( \wz'  \grad \wz - \wz
\grad'\wz'\right),\label{eqn:gamma}
\end{eqnarray}
where we have used Eq. (\ref{eqn:fourier}). This equation can
also be derived directly from the general kinetic equation (\ref{gen7}) as shown
in Appendix B of Ref. \cite{Chavanis2008} [see Eq. (171) of that paper].
\section{Space-Averaged Collision Term\label{sec:space-average}}
Equation (\ref{eqn:gamma}) includes the oscillatory term $\exp(i(\vk+\vk')\cdot(\vr - \vr'))$.
To reveal characteristics of the obtained collision term, 
we need to calculate the space average of the collision term
to drop the high-frequency component.
Space average is calculated over the small rectangular area $\Lambda$ with sides both $2L$
located at $\vec{r}$.
The space average of the diffusion flux $\vec{\Gamma}$ given by Eq. (\ref{eqn:gamma})
is defined by
\begin{equation}
	\langle \vec{\Gamma}\rangle_s \equiv \vec{\Gamma}_s(\vr) = \frac{1}{|\Lambda(\vr)|}\int_{\Lambda(\vr)}d\vr'' \vec{\Gamma}(\vr'').
\end{equation}
We assume that the macroscopic variables such as $\u$ and $\wz$ are
constant inside $\Lambda(\vr)$ so that only the term $\exp(i(\vk+\vk')\cdot(\vr
- \vr'))$
should be space-averaged:
\begin{eqnarray}
	& & \langle \exp(i(\vk+\vk')\cdot(\vr - \vr')) \rangle_s\nonumber\\
	& = & \frac{1}{(2L)^2}\int_{-L}^{L} dx'' \int_{-L}^{L} dy'' \exp(i(\vk + \vk')\cdot \vr'') \exp(-i(\vk+\vk')\cdot (\vr - \vr'))\nonumber\\
	&\approx& \left(\frac{\pi}{L}\right)^2 \delta(\vk + \vk') \exp(-i(\vk+\vk')\cdot(\vr - \vr'))\nonumber\\
	& = & \left(\frac{\pi}{L}\right)^2 \delta(\vk + \vk'),
\end{eqnarray}
where $\vr'' = (x'', y'')$.
Therefore, the space-averaged diffusion flux is given by
\begin{eqnarray}
	& & \vec{\Gamma}_s(\vr)\nonumber\\
	& = & - \Omega \left(\frac{\pi}{L}\right)^2 \int d\vr' \int \frac{d\vk}{(2\pi)^4}
			\pi\delta(\vk \cdot (\u - \u')) \nonumber\\
	& & \times \frac{\hat{\vec{z}} \times \vk}{|\vk|^2} \frac{\hat{\vec{z}} \times \vk}{|\vk|^2} \cdot \left( \wz'  \grad \wz - \wz \grad'\wz'\right). \label{eqn:space_average}
\end{eqnarray}
In Eq. (\ref{eqn:space_average}), we omit the imaginary part as the collision term consists of only the real part.
Further integration over $\vec{k}$ in Eq. (\ref{eqn:space_average}) can be performed.
The integral concerning $\vec{k}$ is as follows:
\begin{equation}
	\int d\vk \delta(\vk \cdot (\u - \u'))
\frac{(\hat{\vec{z}}\times\vk)(\hat{\vec{z}}\times\vk)}{|\vk|^4}
\label{eqn:k-integral}.
\end{equation}
Dividing $\vk$ into the parallel and the perpendicular components and inserting them
into Eq. (\ref{eqn:k-integral}), 
\begin{eqnarray}
	\vk & = & k_{\|} \hat{\vec{n}}_{\|} + k_{\perp}\hat{\vec{n}}_{\perp},\nonumber\\
	\hat{\vec{n}}_{\|} & = & \frac{\u - \u'}{|\u - \u'|}, \nonumber\\
	\hat{\vec{n}}_{\perp} & = & \hat{\vec{z}}\times \hat{\vec{n}}_{\|},
\end{eqnarray}
we obtain
\begin{eqnarray}
	& & \int d\vk \delta(\vk \cdot (\u - \u'))
\frac{(\hat{\vec{z}}\times\vk)(\hat{\vec{z}}\times\vk)}{|\vk|^4} \nonumber\\
	& = & \int d k_{\|} \int d k_{\perp} \delta(k_{\|} |\u - \u'|) \nonumber\\
	& & \qquad \times \frac{ [\hat{\vec{z}} \times (k_{\|} \hat{\vec{n}}_{\|} + k_{\perp} \hat{\vec{n}}_{\perp}) ] [\hat{\vec{z}} \times (k_{\|} \hat{\vec{n}}_{\|} + k_{\perp} \hat{\vec{n}}_{\perp})] }{|k_{\|}^2+k_{\perp}^2|^2}\nonumber\\
	& = & \int d k_{\perp} \frac{1}{|\u - \u'|} \frac{1}{k_{\perp}^4}(\hat{\vec{z}}\times k_{\perp}\hat{\vec{n}}_{\perp}) (\hat{\vec{z}}\times k_{\perp}\hat{\vec{n}}_{\perp})\nonumber\\
	& = & \int d k_{\perp} \frac{1}{|\u - \u'|} \frac{1}{k_{\perp}^2} \frac{\u - \u'}{|\u - \u'|} 
		\frac{\u - \u'}{|\u - \u'|}\nonumber\\
	& = & \frac{(\u - \u') (\u - \u')}{|\u - \u'|^3}2\left[ -\frac{1}{k_{\perp}} \right]_{k_{\rm min}}^{\infty}\nonumber\\
	& = & \frac{(\u - \u') (\u - \u')}{|\u - \u'|^3} \frac{2}{k_{\rm min}},
\end{eqnarray}
where the parameter $k_{\rm min}$ is introduced to regularize a singularity. 
It is determined by the largest wave length that does not exceed the system
size,
namely $k_{\rm min} = 2\pi/R$ where $R$ is a characteristic system size 
determined by an initial distribution of the vortices.

Finally, we obtain the following formulae for the diffusion and drift:
\begin{eqnarray}
	\vec{\Gamma}_s(\vr) & \equiv & -{\sf D}_s(\vr)\cdot \grad\wz + \vec{V}_s(\vr)\wz\nonumber\\
	& = & - K \int d\vr' \frac{(\u - \u')(\u-\u')}{|\u - \u'|^3}\nonumber\\
	& & \cdot (\wz'\grad\wz - \wz\grad'\wz'), \label{space-averaged-gamma-with-new-coeff}\\
	{\sf D}_s & = &  K \int d\vr' \frac{(\u - \u')(\u-\u')\omega'}{|\u - \u'|^3}, \label{eqn:final-D}\\
	\vec{V}_s & = &  K \int d\vr' \frac{(\u - \u')(\u-\u')\cdot \grad'\omega'}{|\u - \u'|^3}, \label{eqn:final-V}\\
		K & \equiv & \frac{\Omega}{(2\pi)^3} \left( \frac{\pi}{L}\right)^2\frac{1}{k_{\rm min}}
		= \frac{\Omega}{R} \left( \frac{R}{4\pi L}\right)^2
\end{eqnarray}
In conclusion, the kinetic equation writes
\begin{eqnarray}
	\frac{\partial {\omega}}{\partial t}+{\vec{u}}\cdot
	\nabla {\omega}=K
	\nabla \cdot \int d{\vr}'\,  \frac{(\u - \u')(\u-\u')}{|\u - \u'|^3} 
	\cdot \left ( {\omega}' \nabla{\omega}- {\omega} \nabla' {\omega}'\right ),
\label{finea}
\end{eqnarray}
and it can be put in the Fokker-Planck form (\ref{fp}).
In the following, we show three good properties of the obtained collision term (\ref{space-averaged-gamma-with-new-coeff}).

\subsection{Collision term in local and global equilibrium states}
At first, let us examine if the collisional effect (\ref{space-averaged-gamma-with-new-coeff}) 
locally disappears in a local equilibrium state.
We rewrite Eq. (\ref{space-averaged-gamma-with-new-coeff}) into a symbolic form:
\begin{equation}
	\vec{\Gamma}_s(\vr) = - K \int d\vr' \vec{\gamma}[\omega, \psi; \omega', \psi'], \label{eqn:functional}
\end{equation}
where $\vec{\gamma}$ is a functional of $\omega$, $\psi$, $\omega'$, and $\psi'$.
Consider a state where the temperature is locally uniform in each small region
in the system. Namely, the whole system consists of subsystems
with different
$\beta$.
We call this state the local equilibrium state
in which the local equilibrium condition is satisfied:
\begin{equation}
	\omega_{\rm leq} = \omega_0\exp(-\beta_{\rm leq}\Omega\psi_{\rm leq}). \label{eqn:equilibrium}
\end{equation}
Inserting Eq. (\ref{eqn:equilibrium}) into $\vec{\gamma}$ in Eq.
(\ref{eqn:functional}), and assuming that ${\vec{r}}$ and
${\vec{r}}'$
belong to the {\it same} subsystem,
we find that
\begin{eqnarray}
	& & \vec{\gamma}[\wz_{\rm leq},\psi_{\rm leq}; \wz'_{\rm leq}, \psi'_{\rm leq}]\nonumber\\
	& = & \frac{(\u_{\rm leq} - \u'_{\rm leq})}{|\u_{\rm leq} - \u'_{\rm leq}|^3}(\u_{\rm leq} - \u'_{\rm leq})\cdot(\wz'_{\rm leq} \grad \wz_{\rm leq} - \wz_{\rm leq} \grad'\wz'_{\rm leq})\nonumber\\
	& = & -\beta_{\rm leq} \Omega\wz_{\rm leq}\wz'_{\rm leq} \frac{(\u_{\rm leq} - \u'_{\rm leq})}{|\u_{\rm leq} - \u'_{\rm leq}|^3}(\u_{\rm leq} - \u'_{\rm leq})\cdot(\grad\psi_{\rm leq} - \grad'\psi'_{\rm leq})\nonumber\\
	& = & 0, \label{eqn:equilibrium2}
\end{eqnarray}
where $\u_{\rm leq} = -\hat{\vec{z}} \times \grad \psi_{\rm leq}$ is used.
As $\u_{\rm leq} - \u'_{\rm leq}$ is perpendicular to $\grad\psi_{\rm leq} - \grad'\psi'_{\rm leq}$,
$\vec{\gamma}$ is equal to zero and this result indicates that a detailed
balance
is achieved.
In this state, the diffusional effect locally disappears but overall $\vec{\Gamma}_s(\vr)$
remains nonzero. On a longer time scale the system finally relaxes to the
thermal equilibrium state but
the relaxation speed is slow.

When the system reaches a global thermal equilibrium state with uniform $\beta$
\cite{Joyce}:
\begin{equation}
	\omega_{\rm eq} = \omega_0\exp(-\beta\Omega\psi_{\rm eq}), \label{eqn:global-equilibrium}
\end{equation}
we obtain
\begin{eqnarray}
	\grad'\wz'_{\rm eq} & = & \wz'_{\rm eq}\frac{\grad'\wz'_{\rm eq}}{\wz'_{\rm eq}}\nonumber\\
	& = & -\beta \Omega \wz'_{\rm eq}(\grad'\psi'_{\rm eq} - \grad \psi_{\rm eq} + \grad \psi_{\rm eq}).
\end{eqnarray}
As $(\u_{\rm eq} - \u'_{\rm eq}) \cdot (\grad'\psi'_{\rm eq} - \grad \psi_{\rm eq}) = 0$,
the drift term in Eq. (\ref{space-averaged-gamma-with-new-coeff}) is rewritten as
\begin{equation}
	\vec{V}_{s,{\rm eq}} = -\beta \Omega {\sf D}_{s,{\rm eq}} \cdot \grad\psi_{\rm eq}
\end{equation}
which is the counterpart of the Einstein relation
\cite{chavanis1998,Chavanis2001}. On the other hand, the
diffusion term writes
\begin{equation}
{{\sf D}_{s,{\rm eq}}} \cdot \nabla\omega_{\rm eq}=-\beta
\Omega\omega_{\rm eq}{\sf D}_{s,{\rm
eq}}\cdot \nabla\psi_{\rm eq},
\end{equation}
so that the total diffusion flux vanishes: 
$\vec{\Gamma}_{s,{\rm eq}}(\vr)\equiv
-{\sf D}_{s,{\rm eq}}\cdot \grad \wz_{\rm eq} + \vec{V}_{s,{\rm eq}}\wz_{\rm
eq}= 0$.
\subsection{Energy-conservative property of collision term}
It is shown that the obtained kinetic equation (\ref{finea}) conserves
the total mean field energy
\begin{eqnarray}
	E & \equiv & \frac{1}{2} \int d\vr \psi \omega\nonumber\\
	& = & \frac{1}{2} \int d\vr \int d\vr' G(\vr-\vr') \omega'\omega.
\end{eqnarray}
Note that the mean field energy $E$ is different from the energy
$\mathscr{H}$
of the point vortex system
\begin{equation}
	\mathscr{H} = -\frac{1}{4\pi}\sum_i \sum_{j \neq i} \Omega_i \Omega_j \ln|\vr_i - \vr_j|.
\end{equation}
Time derivative of the total mean field energy $E$ is given by
\begin{eqnarray}
	\frac{d E}{dt} & = & \frac{1}{2} \int d\vr \int d\vr'G(\vr - \vr')
		\left( \frac{\partial \omega'}{\partial t} \omega + \omega'\frac{\partial \omega}{\partial t}\right) \nonumber\\
	& = & \int d\vr\psi \frac{\partial \omega}{\partial t}. \label{eqn:time-derivative-system-energy}
\end{eqnarray}
Inserting the space-averaged equation of motion
\begin{equation}
	\frac{\partial \omega}{\partial t} + \div(\u\omega) = -\div \vec{\Gamma}_s
\end{equation}
into Eq. (\ref{eqn:time-derivative-system-energy}), we obtain
\begin{eqnarray}
	\frac{d E}{d t} & = & \int d\vr \psi \left( -\div (\u \omega) - \div \vec{\Gamma}_s \right) \nonumber\\
	& = & \int d\vr \grad \psi \cdot \u\omega + \int d\vr \grad \psi \cdot \vec{\Gamma}_s \nonumber\\
	& = & \int d\vr \grad \psi \cdot \vec{\Gamma}_s \nonumber\\
	& = & -K\int d\vr \int d\vr'\grad \psi\cdot \frac{(\u - \u')(\u - \u')}{|\u - \u'|^3} \cdot (\wz'\grad\wz - \wz\grad'\wz'). \label{eqn:dEdt1}
\end{eqnarray}
By permuting the dummy variables $\vr$ and $\vr'$ in Eq. (\ref{eqn:dEdt1})
and taking the half-sum of the resulting expressions, we obtain
\begin{eqnarray}
	\frac{d E}{dt} & = & -\frac{K}{2} \int d\vr \int d\vr'(\grad \psi - \grad'\psi')\cdot \frac{\u - \u'}{|\u - \u'|^3} \nonumber\\
	& & \times (\u - \u')\cdot (\wz'\grad\wz - \wz\grad'\wz')\nonumber\\
	& = & 0.
\end{eqnarray}
We conclude that the obtained collision term conserves the total mean field energy.
\subsection{$H$ theorem}
It is shown that the obtained kinetic equation (\ref{finea}) satisfies an $H$ theorem.
The entropy function $S$ is defined by using the H function:
\begin{eqnarray}
	S & = & -k_B H,\label{eqn:entropy}\\
	H & = & \int d\vr\frac{\omega}{\Omega}\ln \frac{\omega}{\Omega} + {\rm const.}\nonumber\\
	& = & \frac{1}{\Omega}\int d\vr\omega\ln\omega - N\ln \Omega + {\rm
const.}
\end{eqnarray}
The time derivative of the $H$ function is given by
\begin{eqnarray}
	\frac{dH}{dt} & = & \frac{1}{\Omega} \int d\vr  \frac{\partial
\omega}{\partial t} (\ln \omega + 1)\nonumber\\
	& = & \frac{1}{\Omega} \int d\vr (-\div (\u \omega) - \div \vec{\Gamma}_s)(\ln \omega + 1)\nonumber\\
	& = & \frac{1}{\Omega} \int d\vr \u\omega \cdot \grad\ln \omega + \frac{1}{\Omega} \int d\vr \vec{\Gamma}_s \cdot \grad \ln \omega\nonumber\\
	& = & -\frac{1}{\Omega} \int d\vr (\div \u) \omega + \frac{1}{\Omega} \int d\vr \vec{\Gamma}_s \cdot \grad \ln \omega\nonumber\\
	& =& \frac{1}{\Omega} \int d\vr \vec{\Gamma}_s \cdot \grad \ln \omega.\label{eqn:dhdt0}
\end{eqnarray}
Inserting Eq. (\ref{space-averaged-gamma-with-new-coeff}) into Eq. (\ref{eqn:dhdt0}),
we obtain
\begin{eqnarray}
	\frac{dH}{dt} & = & -\frac{K}{\Omega} \int d\vr \int d\vr' \frac{\grad\wz}{\wz} \cdot \frac{(\u-\u')(\u-\u')}{|\u-\u'|^3}\nonumber\\
	& & \cdot(\wz'\grad\wz - \wz\grad'\wz'). \label{eqn:dHdt1}
\end{eqnarray}
By permuting the dummy variables $\vr$ and $\vr'$ in Eq. (\ref{eqn:dHdt1})
and taking the half-sum of the resulting expressions, we obtain
\begin{eqnarray}
	\frac{dH}{dt} & = & - \frac{K}{2\Omega} \int d\vr \int d\vr' \frac{1}{\wz\wz'}(\wz'\grad\wz - \wz\grad'\wz') \cdot \frac{\u-\u'}{|\u-\u'|^3}\nonumber\\
	&   & \times (\u - \u')\cdot(\wz'\grad\wz - \wz\grad'\wz')  \nonumber\\
	& = & -\frac{K}{2 \Omega} \int d\vr \int d\vr' \frac{1}{\wz\wz'}\frac{|(\u-\u')\cdot(\wz'\grad\wz - \wz\grad'\wz')|^2}{|\u-\u'|^3} \nonumber\\
	& \le & 0.\label{eqn:dHdt2}
\end{eqnarray}
The integrand of Eq. (\ref{eqn:dHdt2}) is positive or equal to zero,
and $dH/dt$ is negative or equal to zero.
It is concluded that the entropy function $S$ defined by Eq. (\ref{eqn:entropy})
is a monotonically increasing function. This ensures that 
the system reaches the Boltzmann equilibrium state (\ref{eqn:global-equilibrium}) in the macroscopic fluid scale.
\section{Discussion\label{sec:discussion}}
We have derived a kinetic equation  of the 
Fokker-Planck type for point vortices [see Eq. (\ref{finea})].
The collision term exhibits several important properties:
(a) it includes the nonlocal, long-range, interaction;
(b) it conserves the mean field energy; 
(c) it satisfies the $H$ theorem;
(d) its effect vanishes in each local equilibrium region with the same temperature.
When the system reaches a global equilibrium state, the collision term completely
converges to zero all over the system. 

The order of the obtained diffusion flux
(\ref{space-averaged-gamma-with-new-coeff}) is $O(\epsilon)$.
On the other hand, simple calculation shows that the term is proportional
to $1/N$. Thus, it is found that the expansion parameter
$\epsilon$ is of the order $1/{N}$.

The kinetic equation
(\ref{finea})
structurally differs
from the previously obtained kinetic equation (\ref{grd}) because in Eq.
(\ref{finea}) the conservation of energy is
ensured by the tensor ${(\u - \u')(\u-\u')}/{|\u - \u'|^3}$ while in Eq.
(\ref{grd}) the conservation of energy is
ensured by the delta function  $\delta({\vec{x}}\cdot
{\vec{w}})$ accounting for a
condition of resonance (the
tensor ${{\vec{x}}_{\perp}\otimes
{\vec{x}}_{\perp}}/{x^2}$ in Eq. (\ref{grd})
ensures the conservation of angular momentum). Therefore, these kinetic
equations cannot be reconciled and they have, at best, a different domain of
validity. Equation (\ref{grd}) is expected to be valid when the mean flow is
strong (for axisymmetric flows,  Eq. (\ref{sd4}) can be derived rigoroulsy at
the order $1/N$ when $N\rightarrow +\infty$) while Eq. (\ref{finea}) is expected
to be valid when the mean flow is weak. This may be the case
when $N$ is ``not too large'' so that the fluctuations are important (see
Appendix \ref{sec_val}). However, since this equation cannot be derived in a
well-defined
mathematical limit, some problems arise that we briefly discuss.

(i) The final formulae (\ref{eqn:final-D}) and
(\ref{eqn:final-V}) include unknown parameters $k_{\rm min}$ and $L$. These
cut-offs arise because the assumptions made to derive the kinetic equation do
not correspond to a well-defined asymptotic limit (e.g. $N$ must be large but
not too much). Therefore, our approach must be considered as being heuristic and
some cut-offs must be introduced by hand (or adapted to the situation).

(ii) The integrals in Eqs. (\ref{eqn:final-D}) for ${\sf
D}$
and (\ref{eqn:final-V}) for $\vec{V}$ diverge individually. This is a problem
if we consider a test particle approach. However, this is not a problem to
describe the evolution of the system as a whole since the combined term
$\vec{\Gamma}_s = -{\sf D}_s(\vr)\cdot \grad\wz + \vec{V}_s(\vr)\wz$
 converges [see also point (iv)].

(iii) The kinetic equation (\ref{finea})  does not conserve
the 
angular momentum even in an infinite domain or in a circular domain (contrary
to the kinetic equation (\ref{grd})). This may be 
related to our assumption that the mean field is weak so that the system does
not ``see'' the symmetries of the system. Actually, the same kinetic equation
would be obtained in a bounded domain with only a change in the parameters 
$k_{\rm min}$ and $L$
that are unknown anyway.

(iv) The diffusion flux takes large values when
${\vec{u}}({\vec{r}}')\rightarrow
{\vec{u}}({\vec{r}})$ with
${\vec{r}}'\neq {\vec{r}}$.
This feature is problematic on a
physical point of view because it implies that the
interaction between two point vortices that are far away but that have,
coincidentally, the same
velocity contributes importantly to the diffusion flux. One
would expect, on
the contrary, that the contribution of far away vortices decreases with the
distance. Indeed, point vortices do not
``see'' each
other if they are far away. This bad feature adds to the other
divergences
mentioned in points (i) and (ii) above. However, the precise form of the
collision kernel  ${(\u - \u')(\u-\u')}/{|\u - \u'|^3}$ is not of
main importance. What really matters is that it is proportional to the tensor 
${(\u - \u')(\u-\u')}$ 
in order to satisfy the conservation of the
energy \cite{Note04}
and the other nice
properties
discussed in Sec. \ref{sec:space-average}. The function
$\chi=1/{|\u - \u'|^3}$
in factor of this
tensor could be changed in
order to avoid un-physical divergences or undesirable features, while keeping
the main properties of the kinetic equation.

Despite all these limitations, we think that the present 
approach has some interest since it leads to an explicit kinetic equation
(\ref{finea}) that
could  be solved numerically and confronted to direct numerical simulations.
Surely, the next step would be to test numerically the relevance of this kinetic
equation  and
determine the parameters $L$, $k_{\rm min}$ and
$\chi$ that are
ill-defined or pose problem.

\section{Comparison with Numerical Simulations \label{sec:compns}}

Finally, we would like to compare the predictions of statistical
mechanics and
kinetic theory of 2D point vortices with numerical simulations.

\subsection{Previous simulations \label{sec:kn}}

A very interesting numerical work has been performed by Kawahara and Nakanishi
\cite{kn}. 
Using different types of initial conditions, they observe that the system
settles 
down to a final state via a slow collisional relaxation after relaxing into a
quasi
stationary state via an initial violent collisionless relaxation. 
Their numerical results show that (i) the Boltzmann entropy increases
monotonically, 
(ii) the system relaxes towards the maximum entropy state (statistical
equilibrium state), 
(iii) the relaxation time increases linearly with the number $N$ of point
vortices.

The two-stages relaxation process consisting of the violent
and slow relaxations confirms the theoretical predictions made 
earlier by Chavanis \cite{Chavanis2002,Chavanis2001}  using an analogy with the
dynamics of stellar 
systems.
Although the initial condition in Ref. \cite{kn} is axisymmetric, the slow
collisional 
evolution is non-axisymmetric (see their Fig. 7). 
Actually, during the violent relaxation stage, the system forms 
macroscopic clusters that are called ``vortex crystals''
\cite{Fine1995}. 
They are steady states of the 2D Euler equation describing the collisionless
regime. 
For $N\rightarrow +\infty$, these clusters would persist for all times. 
The kinetic theory explains how they are slowly destroyed by finite $N$ effects
and 
point vortex ``collisions''. 
As a result, they finally disappear through successive mergings
as observed in Ref. \cite{kn}.

The obtained kinetic equation (\ref{finea}) explains how the non-axisymmetric
profile
after the violent relaxation evolves towards the Boltzmann distribution
following the $H$ theorem in Eq. (\ref{eqn:dHdt2}).
As the order of the obtained collision term
(\ref{space-averaged-gamma-with-new-coeff}) is $O(N^{-1})$,
we expect that the relaxation time of the non-axisymmetic profile scales 
as $N$.
This is precisely what Kawahara and Nakanishi observe
numerically \cite{Note05}.

By contrast, for a purely axisymmetric evolution, we expect a very different 
behavior since the general kinetic equation (\ref{gen7}) simplifies in Eq.
(\ref{sd4}). 
This equation does not relax towards the Boltzmann distribution and the 
collision term even reduces to zero when the profile of angular velocity is 
monotonic, implying a relaxation time scaling as $N^2$ or being even longer 
(see the discussion in Refs. \cite{Chavanis2012,Chavanis2012b,cl,Chavanis2010}).
These considerations show that the dynamics of 
point vortices is very complex and that different regimes, described
by different kinetic theories, may occur. It would be interesting to study
numerically these different regimes in future works.

\subsection{A new set of numerical simulations\label{sec:newsim}}

In order to illustrate our theoretical study, we have performed a new set of
numerical simulations. This numerical work is preliminary and a more detailed
study will be the subject of a specific paper.

We wish to test some general properties of the kinetic theory, notably the
$H$-theorem and the relaxation time. The characteristic time scale of the
relaxation $t_R$ is determined by the kinetic equation (\ref{finea}); it
scales as $t_R\propto 1/\Omega$ and is independent of $N$.
On the other hand, the characteristic dynamical time $t_D$ is determined by
a rotation time of a circular clump distribution of the vortices; it scales as
$t_D\propto 1/(N\Omega)$.
In the following, the scalings $t_R \propto 1/\Omega$ and $t_R/t_D \propto N$
are demonstrated numerically.

\begin{figure}
\resizebox{3cm}{!}{\includegraphics{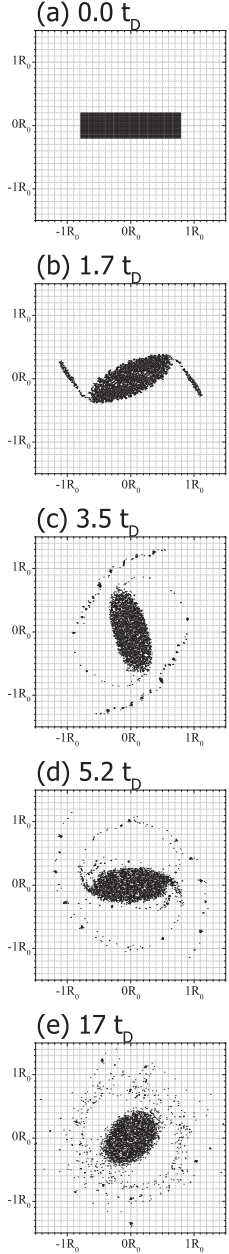}}
\caption{A typical time evolution of the vortices with $\Omega=0.15$, $N=3751$ is shown.
Cell size used for the coarse-graining is indicated by the grid lines in gray.}
\label{fig:time-evolution}
\end{figure}

\begin{figure}
\resizebox{7cm}{!}{\includegraphics{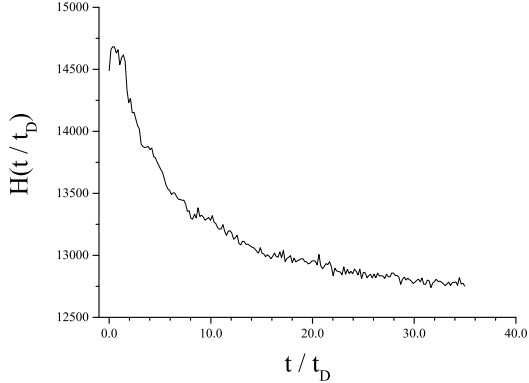}}
\caption{The time evolution of the H function with $\Omega=0.15$, $N=3751$ is
shown.
}
\label{fig:h-function}
\end{figure}

We consider a single-species point vortex system in an infinite domain.
Point vortices are arranged in a rectangular area with sides $1.6 R_0$ by $0.4
R_0$
where $R_0$ is a macroscopic characteristic length of the system.
As the obtained kinetic equation (\ref{finea}) explains the
relaxation process
of a non-axisymmetric profile, we choose the initial profile as
a non-axisymmetric one.
A typical time evolution of the vortices is shown in Fig. \ref{fig:time-evolution}.
The H function is numerically determined by 
\begin{eqnarray}
	H(t)    =  \sum_i (\Delta X)^2 n_i(t) \ln n_i(t),
\end{eqnarray}
with
\begin{eqnarray}
	n_i(t)  =  \frac{N_i(t)}{(\Delta X)^2},	\qquad\Delta X  = 
\frac{R_0}{10}.
\end{eqnarray}
The configuration space is divided into square cells with side $\Delta X$
as is shown by the grid lines in gray in Fig. \ref{fig:time-evolution}.
The number of  vortices in the $i$th cell is denoted $N_i(t)$.
The time evolution of the H function corresponding to Fig.
\ref{fig:time-evolution}
is shown in Fig. \ref{fig:h-function}.
As is predicted by the H theorem, the value of the H function monotonically 
decreases, which assures that the system settles down to a final equilibrium state 
characterized by the maximum entropy.

\begin{figure}
\resizebox{7cm}{!}{\includegraphics{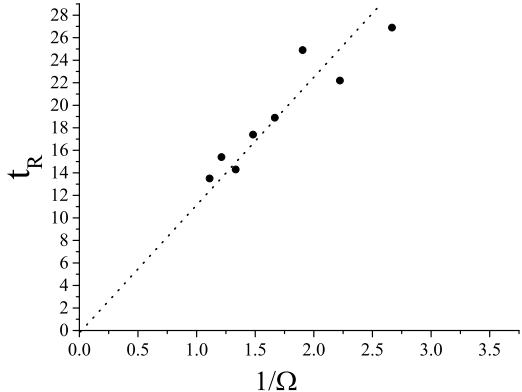}}
\caption{The relaxation time $t_R$ is plotted against $\Omega^{-1}$ for fixed
$N=3751$. 
It indicates that $t_R$ is proportional to $\Omega^{-1}$.}
\label{fig:tr-vs-inv-omega}
\end{figure}

\begin{figure}
\resizebox{7cm}{!}{\includegraphics{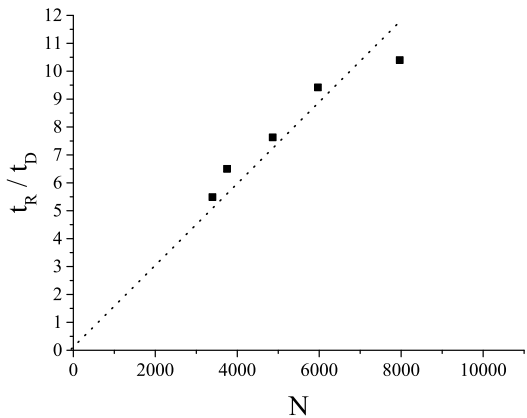}}
\caption{The relaxation time $t_R$ scaled by the dynamical time $t_D$
is plotted against $N$ for fixed $\Omega=0.15$. 
It indicates that $t_R/t_D$ is proportional to $N$.}
\label{fig:tr-vs-n}
\end{figure}

We assume a temporal evolution of the H function as
\begin{equation}
	H(t) = H_0 + H_1 \exp\left( -\frac{t}{t_R} \right).
\end{equation}
To determine the relaxation time numerically, simulations with
(a) fixed $\Omega$, variable $N$ and (b) fixed $N$, variable $\Omega$
are carried out. The plots of $t_R$ versus $1/\Omega$ and $t_R/t_D$
versus $N$ are shown in Figs.  \ref{fig:tr-vs-inv-omega} and \ref{fig:tr-vs-n}.
These plots elucidate the relations $t_R \propto 1/\Omega$ and
$t_R/t_D \propto N$.

The scaling $t_R/t_D \propto N$ for a non-axisymmetric flow agrees with the one 
previously obtained by Kawahara and Nakanishi  \cite{kn}. This
agreement is interesting because the numerical conditions are different.
We work in an unbounded domain and start from a rectangular patch while they
work in a bounded domain and start from an annulus. Since our domain is
unbounded, no vortex crystal forms. Therefore, the evolution is different but
the scaling  $t_R/t_D \propto N$ is the same. Furthermore, we obtain this
scaling with a larger number of vortices. Finally, we numerically demonstrate a
new scaling law  $t_R \propto 1/\Omega$ which is predicted by the kinetic
equation (\ref{finea}). These numerical results agree with the kinetic theory
developed in this paper. It would be nevertheless useful to ascertain these
results by making longer simulations and study in more detail the distinction
between the collisionless
regime (violent relaxation) and the collisional one (slow relaxation). This
will be the object of a future paper.


\section{Conclusion\label{sec:conclusion}}
We have developed a kinetic theory of point vortices for weak mean flows for which the 
timescale of the macroscopic motion is long as compared to the decorrelation time. A linear trajectory approximation
has been used as a first order approximation to compute the correlation functions appearing in the collision term
of the kinetic equation. The smooth vorticity field is the solution of the kinetic equation (\ref{finea}). 
The results of the kinetic theory agree with the numerical simulations: (i) the 
H function decreases monotonically ($H$ theorem); (ii) the system
relaxes towards the Boltzmann distribution; (iii) the relaxation time
scales as $t_R \propto 1/\Omega$ and $t_R/t_D \propto N$.
The kinetic theory of point vortices describing ``collisions'' between point
vortices and finite $N$ effects can explain, for example,
the destruction of ``vortex
crystals'' that form during the collisionless regime \cite{Fine1995,kn}. 
Therefore, the kinetic theory developed in the
present paper can account for many results of numerical simulations and
laboratory experiments. There remains open  issues such as the difference of
behavior between axisymmetric and non-axisymmetric evolutions that will be 
considered in future works.

\begin{acknowledgments}
This work was supported by JSPS KAKENHI Grant Number 24540400.
\end{acknowledgments}

\appendix

\section{Validity of Linear Trajectory Approximation}
\label{sec_val}
The validity of the linear trajectory approximation is assessed
as follows.
There are two important characteristic time scales.
One corresponds to the Brownian motion and will be 
denoted by $\tau_B$.
The other corresponds to the macroscopic 
fluid motion and will be denoted by $\tau_F$.
Using these two characteristic time scales, we will 
derive a condition where the linear trajectory approximation
may be valid.

Consider a microscopic area with both sides $2L$, which
is of the same size as the space-averaging area introduced 
in Sec. \ref{sec:space-average}.
The box moves with the macroscopic flow velocity.
Inside the box, vortices fluctuate due to the Brownian motion.
Vortices stay within the same box if the time period is short
and the macroscopic orbit of a point vortex may be along the
flow.
However, if the time period exceeds a certain value, say $\tau_B$,
vortices leave the box and the macroscopic fluid approximation is
no longer valid.
The stochastic process due to the Brownian motion is expressed 
by Eq. (\ref{eqn:cumulant}) and $\tau_B$ is estimated by
\begin{equation}
	D \tau_B = (2L)^2. \label{eqn:a1}
\end{equation}
Within the time scale shorter than $\tau_B$, the macroscopic fluid
approximation is valid.

On the other hand, the condition that the linear trajectory
approximation is valid 
is equivalent to the time scale where the macroscopic 
flow orbit is straight.
We assume that the macroscopic flow has a circular orbit with a
radius of curvature $R_0$.
As the scale length concerned is macroscopic and longer than $2L$, the Brownian
motion
cannot be seen.
Then the characteristic time scale $\tau_F$ which corresponds
to the macroscopic fluid motion may be estimated by
\begin{equation}
	u \tau_F = 2\pi R_0 \label{eqn:a2}
\end{equation}
which is a turnover time with uniform velocity $u$.
In our model, the flow orbit consists of many fragment
orbits.
Each fragment orbit is approximately obtained by the linear
trajectory
approximation.
The time for a vortex to cross a fragment orbit is less than 
$\tau_B$.

To approximate the integral as shown in Eq. (\ref{eqn:defofnu}), $\nu (t-t_0)$
must be smaller than 1, namely,
\begin{equation}
	\frac{1}{\nu} < t-t_0.
\end{equation}
By definition, the order of $1/\nu$ is estimated by
\begin{equation}
	\frac{1}{\nu} = \frac{2}{k^2D} = \frac{2}{D} \frac{(2L)^2}{(2\pi)^2} = \frac{2}{(2\pi)^2}\tau_B,
\end{equation}
where we have used the relation
\begin{equation}
	k = \frac{2\pi}{2L}.
\end{equation}

>From Eq. (\ref{eqn:final-D}), the order of $D$ is
estimated by
\begin{eqnarray}
	D & = & \frac{\Omega}{(2\pi)^3} \left( \frac{\pi}{L}\right)^2 \frac{1}{k_{\rm min}} \int \cdots\nonumber\\
	& \propto & \frac{\Omega}{(2\pi)^3} \left( \frac{\pi}{L}\right)^2 \frac{1}{k_{\rm min}} \frac{N\Omega}{u} \nonumber\\
	& = & \frac{1}{N}\frac{1}{(2\pi)^3} \left( \frac{\pi}{L}\right)^2 \frac{1}{k_{\rm min}} \frac{N\Omega}{u}.\label{eqn:a6}
\end{eqnarray}
The circular flow speed is approximated by its average value as
\begin{equation}
	u = R_0\omega = R_0 \frac{N\Omega}{\pi R^2} = \frac{N\Omega}{\pi} \frac{R_0}{R^2} \label{eqn:a7}
\end{equation}
and
\begin{equation}
	\pi R^2\omega = N\Omega, \label{eqn:a8}
\end{equation}
where $R$ is a characteristic length of the system ($R > R_0$).
Using Eqs. (\ref{eqn:a1}), (\ref{eqn:a2}), (\ref{eqn:a6}), (\ref{eqn:a7}), (\ref{eqn:a8}) and the relation 
$k_{\rm min} = 2\pi/R$, 
inequality $\tau_B < \tau_F$ is rewritten as
\begin{equation}
	(2L)^2\left( \frac{1}{N} \frac{1}{(2\pi)^3} \frac{\pi^2}{L^2} \frac{R}{2\pi} \frac{N\Omega}{u}\right)^{-1} < \frac{2\pi R_0}{u}
\end{equation}
or as
\begin{equation}
	N < \frac{\pi}{32} \frac{1}{N\Omega} \frac{1}{L^4} \frac{R^5}{R_0}. \label{eqn:a10}
\end{equation}
Equation (\ref{eqn:a10}) gives an upper limit on $N$.

On the other hand, in the usual hydrodynamic picture, 
there is a lower limit length scale $L_0$.
A region with size $L_0 \times L_0$
contains only one particle and a larger region with size $L \times L$
contains enough particles and reaches a local thermal equilibrium state. Therefore,
\begin{equation}
	L_0 < L < R.
\end{equation}
Then the ratio of the number of the particles is given by
\begin{equation}
	\frac{L_0^2}{\pi R^2} = \frac{1}{N}
\end{equation}
which implies
\begin{equation}
	\frac{1}{N} = \frac{1}{\pi} \frac{L_0^2}{R^2} < \frac{1}{\pi}\frac{L^2}{R^2}.
\end{equation}
Namely, the following relation is obtained:
\begin{equation}
	1 < \pi \left( \frac{R}{L}\right)^2 < N. \label{eqn:a14}
\end{equation}
It is found that there is a lower limit on $N$.

Using Eqs. (\ref{eqn:a10}) and (\ref{eqn:a14}), we
obtain the condition that 
the linear trajectory approximation is valid:
\begin{equation}
	1 < \pi \left( \frac{R}{L}\right)^2 < N < \frac{\pi}{32} \frac{1}{N\Omega} \frac{1}{L^4} \frac{R^5}{R_0}. \label{eqn:a15}
\end{equation}
If we set $R_0 = R/2$ and $N\Omega=1$, Eq. (\ref{eqn:a15}) is rewritten as
\begin{equation}
	1 < \pi \left( \frac{R}{L}\right)^2 < N < \frac{\pi}{16} \left(\frac{R}{L}\right)^4.
\end{equation}
As we used an $1/N$ expansion to evaluate the collision term,
$N$ must be large.
But the above evaluation indicates that the value of $N$ has a certain maximum
number
to validate the linear trajectory approximation.
The estimates given above are very approximate but their main goal
 is to suggest that there is a non-asymptotic regime in which our approach applies.


\begin{thebibliography}{45}%
\makeatletter
\providecommand \@ifxundefined [1]{%
 \@ifx{#1\undefined}
}%
\providecommand \@ifnum [1]{%
 \ifnum #1\expandafter \@firstoftwo
 \else \expandafter \@secondoftwo
 \fi
}%
\providecommand \@ifx [1]{%
 \ifx #1\expandafter \@firstoftwo
 \else \expandafter \@secondoftwo
 \fi
}%
\providecommand \natexlab [1]{#1}%
\providecommand \enquote  [1]{``#1''}%
\providecommand \bibnamefont  [1]{#1}%
\providecommand \bibfnamefont [1]{#1}%
\providecommand \citenamefont [1]{#1}%
\providecommand \href@noop [0]{\@secondoftwo}%
\providecommand \href [0]{\begingroup \@sanitize@url \@href}%
\providecommand \@href[1]{\@@startlink{#1}\@@href}%
\providecommand \@@href[1]{\endgroup#1\@@endlink}%
\providecommand \@sanitize@url [0]{\catcode `\\12\catcode `\$12\catcode
  `\&12\catcode `\#12\catcode `\^12\catcode `\_12\catcode `\%12\relax}%
\providecommand \@@startlink[1]{}%
\providecommand \@@endlink[0]{}%
\providecommand \url  [0]{\begingroup\@sanitize@url \@url }%
\providecommand \@url [1]{\endgroup\@href {#1}{\urlprefix }}%
\providecommand \urlprefix  [0]{URL }%
\providecommand \Eprint [0]{\href }%
\providecommand \doibase [0]{http://dx.doi.org/}%
\providecommand \selectlanguage [0]{\@gobble}%
\providecommand \bibinfo  [0]{\@secondoftwo}%
\providecommand \bibfield  [0]{\@secondoftwo}%
\providecommand \translation [1]{[#1]}%
\providecommand \BibitemOpen [0]{}%
\providecommand \bibitemStop [0]{}%
\providecommand \bibitemNoStop [0]{.\EOS\space}%
\providecommand \EOS [0]{\spacefactor3000\relax}%
\providecommand \BibitemShut  [1]{\csname bibitem#1\endcsname}%
\let\auto@bib@innerbib\@empty
\bibitem [{\citenamefont {Newton}(2001)}]{Newton}%
  \BibitemOpen
  \bibfield  {author} {\bibinfo {author} {\bibfnamefont {P.}~\bibnamefont
  {Newton}},\ }\href@noop {} {\emph {\bibinfo {title} {The N-Vortex Problem:
  Analytical Techniques}}}\ (\bibinfo  {publisher} {Springer-Verlag},\ \bibinfo
  {address} {Berlin},\ \bibinfo {year} {2001})\BibitemShut {NoStop}%
\bibitem [{\citenamefont {Kraichnan}\ and\ \citenamefont
  {Montgomery}(1980)}]{Kraichnan}%
  \BibitemOpen
  \bibfield  {author} {\bibinfo {author} {\bibfnamefont {R.~H.}\ \bibnamefont
  {Kraichnan}}\ and\ \bibinfo {author} {\bibfnamefont {D.}~\bibnamefont
  {Montgomery}},\ }\href@noop {} {\bibfield  {journal} {\bibinfo  {journal}
  {Rep. Prog. Phys.}\ }\textbf {\bibinfo {volume} {43}},\ \bibinfo {pages}
  {547} (\bibinfo {year} {1980})}\BibitemShut {NoStop}%
\bibitem [{\citenamefont {Tabeling}(2002)}]{Tabeling}%
  \BibitemOpen
  \bibfield  {author} {\bibinfo {author} {\bibfnamefont {P.}~\bibnamefont
  {Tabeling}},\ }\href@noop {} {\bibfield  {journal} {\bibinfo  {journal}
  {Phys. Rep.}\ }\textbf {\bibinfo {volume} {362}},\ \bibinfo {pages} {1}
  (\bibinfo {year} {2002})}\BibitemShut {NoStop}%
\bibitem [{\citenamefont {Onsager}(1949)}]{Onsager}%
  \BibitemOpen
  \bibfield  {author} {\bibinfo {author} {\bibfnamefont {L.}~\bibnamefont
  {Onsager}},\ }\href@noop {} {\bibfield  {journal} {\bibinfo  {journal} {Nuovo
  Cimento Suppl.}\ }\textbf {\bibinfo {volume} {6}},\ \bibinfo {pages} {279}
  (\bibinfo {year} {1949})}\BibitemShut {NoStop}%
\bibitem [{\citenamefont {Eyink}\ and\ \citenamefont
  {Sreenivasan}(2006)}]{Eyink}%
  \BibitemOpen
  \bibfield  {author} {\bibinfo {author} {\bibfnamefont {G.~L.}\ \bibnamefont
  {Eyink}}\ and\ \bibinfo {author} {\bibfnamefont {K.~R.}\ \bibnamefont
  {Sreenivasan}},\ }\href@noop {} {\bibfield  {journal} {\bibinfo  {journal}
  {Rev. Mod. Phys.}\ }\textbf {\bibinfo {volume} {78}},\ \bibinfo {pages} {87}
  (\bibinfo {year} {2006})}\BibitemShut {NoStop}%
\bibitem [{\citenamefont {Joyce}\ and\ \citenamefont
  {Montgomery}(1973)}]{Joyce}%
  \BibitemOpen
  \bibfield  {author} {\bibinfo {author} {\bibfnamefont {G.}~\bibnamefont
  {Joyce}}\ and\ \bibinfo {author} {\bibfnamefont {D.}~\bibnamefont
  {Montgomery}},\ }\href@noop {} {\bibfield  {journal} {\bibinfo  {journal} {J.
  Plasma Phys.}\ }\textbf {\bibinfo {volume} {10}},\ \bibinfo {pages} {107}
  (\bibinfo {year} {1973})}\BibitemShut {NoStop}%
\bibitem [{\citenamefont {Montgomery}\ and\ \citenamefont
  {Joyce}(1974)}]{Montgomery1974}%
  \BibitemOpen
  \bibfield  {author} {\bibinfo {author} {\bibfnamefont {D.}~\bibnamefont
  {Montgomery}}\ and\ \bibinfo {author} {\bibfnamefont {G.}~\bibnamefont
  {Joyce}},\ }\href@noop {} {\bibfield  {journal} {\bibinfo  {journal} {Phys.
  Fluids}\ }\textbf {\bibinfo {volume} {17}},\ \bibinfo {pages} {1139}
  (\bibinfo {year} {1974})}\BibitemShut {NoStop}%
\bibitem [{\citenamefont {Kida}(1975)}]{Kida1975}%
  \BibitemOpen
  \bibfield  {author} {\bibinfo {author} {\bibfnamefont {S.}~\bibnamefont
  {Kida}},\ }\href@noop {} {\bibfield  {journal} {\bibinfo  {journal} {J. Phys.
  Soc. Jpn.}\ }\textbf {\bibinfo {volume} {39}},\ \bibinfo {pages} {1395}
  (\bibinfo {year} {1975})}\BibitemShut {NoStop}%
\bibitem [{\citenamefont {Kraichnan}(1975)}]{Kraichnan1975}%
  \BibitemOpen
  \bibfield  {author} {\bibinfo {author} {\bibfnamefont {R.~H.}\ \bibnamefont
  {Kraichnan}},\ }\href@noop {} {\bibfield  {journal} {\bibinfo  {journal} {J.
  Fluid Mech.}\ }\textbf {\bibinfo {volume} {67}},\ \bibinfo {pages} {155}
  (\bibinfo {year} {1975})}\BibitemShut {NoStop}%
\bibitem [{\citenamefont {Seyler{, Jr.}}(1976)}]{Seyler1976}%
  \BibitemOpen
  \bibfield  {author} {\bibinfo {author} {\bibfnamefont {C.~E.}\ \bibnamefont
  {Seyler{, Jr.}}},\ }\href@noop {} {\bibfield  {journal} {\bibinfo  {journal}
  {Phys. Fluids}\ }\textbf {\bibinfo {volume} {19}},\ \bibinfo {pages} {1336}
  (\bibinfo {year} {1976})}\BibitemShut {NoStop}%
\bibitem [{\citenamefont {Pointin}\ and\ \citenamefont
  {Lundgren}(1976)}]{Pointin1976}%
  \BibitemOpen
  \bibfield  {author} {\bibinfo {author} {\bibfnamefont {Y.~B.}\ \bibnamefont
  {Pointin}}\ and\ \bibinfo {author} {\bibfnamefont {T.~S.}\ \bibnamefont
  {Lundgren}},\ }\href@noop {} {\bibfield  {journal} {\bibinfo  {journal}
  {Phys. Fluids}\ }\textbf {\bibinfo {volume} {19}},\ \bibinfo {pages} {1459}
  (\bibinfo {year} {1976})}\BibitemShut {NoStop}%
\bibitem [{\citenamefont {Lundgren}\ and\ \citenamefont
  {Pointin}(1977)}]{Lundgren1977}%
  \BibitemOpen
  \bibfield  {author} {\bibinfo {author} {\bibfnamefont {T.~S.}\ \bibnamefont
  {Lundgren}}\ and\ \bibinfo {author} {\bibfnamefont {Y.~B.}\ \bibnamefont
  {Pointin}},\ }\href@noop {} {\bibfield  {journal} {\bibinfo  {journal} {J.
  Stat. Phys.}\ }\textbf {\bibinfo {volume} {17}},\ \bibinfo {pages} {323}
  (\bibinfo {year} {1977})}\BibitemShut {NoStop}%
\bibitem [{\citenamefont {Ting}\ \emph {et~al.}(1987)\citenamefont {Ting},
  \citenamefont {Chen},\ and\ \citenamefont {Lee}}]{Ting1987}%
  \BibitemOpen
  \bibfield  {author} {\bibinfo {author} {\bibfnamefont {A.~C.}\ \bibnamefont
  {Ting}}, \bibinfo {author} {\bibfnamefont {H.~H.}\ \bibnamefont {Chen}}, \
  and\ \bibinfo {author} {\bibfnamefont {Y.~C.}\ \bibnamefont {Lee}},\
  }\href@noop {} {\bibfield  {journal} {\bibinfo  {journal} {Physica}\ }\textbf
  {\bibinfo {volume} {26D}},\ \bibinfo {pages} {37} (\bibinfo {year}
  {1987})}\BibitemShut {NoStop}%
\bibitem [{\citenamefont {Robert}\ and\ \citenamefont
  {Sommeria}(1991)}]{Robert1991}%
  \BibitemOpen
  \bibfield  {author} {\bibinfo {author} {\bibfnamefont {R.}~\bibnamefont
  {Robert}}\ and\ \bibinfo {author} {\bibfnamefont {J.}~\bibnamefont
  {Sommeria}},\ }\href@noop {} {\bibfield  {journal} {\bibinfo  {journal} {J.
  Fluid Mech.}\ }\textbf {\bibinfo {volume} {229}},\ \bibinfo {pages} {291}
  (\bibinfo {year} {1991})}\BibitemShut {NoStop}%
\bibitem [{\citenamefont {Eyink}\ and\ \citenamefont
  {Spohn}(1993)}]{Eyink1993}%
  \BibitemOpen
  \bibfield  {author} {\bibinfo {author} {\bibfnamefont {G.~L.}\ \bibnamefont
  {Eyink}}\ and\ \bibinfo {author} {\bibfnamefont {H.}~\bibnamefont {Spohn}},\
  }\href@noop {} {\bibfield  {journal} {\bibinfo  {journal} {J. Stat. Phys.}\
  }\textbf {\bibinfo {volume} {70}},\ \bibinfo {pages} {833} (\bibinfo {year}
  {1993})}\BibitemShut {NoStop}%
\bibitem [{\citenamefont {B{\"u}hler}(2002)}]{Buehler2002}%
  \BibitemOpen
  \bibfield  {author} {\bibinfo {author} {\bibfnamefont {O.}~\bibnamefont
  {B{\"u}hler}},\ }\href@noop {} {\bibfield  {journal} {\bibinfo  {journal}
  {Phys. Fluids}\ }\textbf {\bibinfo {volume} {14}},\ \bibinfo {pages} {2139}
  (\bibinfo {year} {2002})}\BibitemShut {NoStop}%
\bibitem [{\citenamefont {Yatsuyanagi}\ \emph {et~al.}(2005)\citenamefont
  {Yatsuyanagi}, \citenamefont {Kiwamoto}, \citenamefont {Tomita},
  \citenamefont {Sano}, \citenamefont {Yoshida},\ and\ \citenamefont
  {Ebisuzaki}}]{Yatsuyanagi2005}%
  \BibitemOpen
  \bibfield  {author} {\bibinfo {author} {\bibfnamefont {Y.}~\bibnamefont
  {Yatsuyanagi}}, \bibinfo {author} {\bibfnamefont {Y.}~\bibnamefont
  {Kiwamoto}}, \bibinfo {author} {\bibfnamefont {H.}~\bibnamefont {Tomita}},
  \bibinfo {author} {\bibfnamefont {M.~M.}\ \bibnamefont {Sano}}, \bibinfo
  {author} {\bibfnamefont {T.}~\bibnamefont {Yoshida}}, \ and\ \bibinfo
  {author} {\bibfnamefont {T.}~\bibnamefont {Ebisuzaki}},\ }\href@noop {}
  {\bibfield  {journal} {\bibinfo  {journal} {Phys. Rev. Lett.}\ }\textbf
  {\bibinfo {volume} {94}},\ \bibinfo {pages} {054502} (\bibinfo {year}
  {2005})}\BibitemShut {NoStop}%
\bibitem [{\citenamefont {Chavanis}(2002)}]{Chavanis2002}%
P.~H. Chavanis, {\it Statistical mechanics of two-dimensional
vortices and stellar systems} in {\it Dynamics and thermodynamics of
systems with long range interactions}, edited by T. Dauxois {\it et al.},
Lecture Notes in Physics {\bf 602}, 208 (Springer, 2002)
\bibitem [{\citenamefont {Matthaeus}\ \emph {et~al.}(1991)\citenamefont
  {Matthaeus}, \citenamefont {Stribling}, \citenamefont {Martinez},
  \citenamefont {Oughton},\ and\ \citenamefont {Montgomery}}]{Matthaeus1991}%
  \BibitemOpen
  \bibfield  {author} {\bibinfo {author} {\bibfnamefont {W.~H.}\ \bibnamefont
  {Matthaeus}}, \bibinfo {author} {\bibfnamefont {W.~T.}\ \bibnamefont
  {Stribling}}, \bibinfo {author} {\bibfnamefont {D.}~\bibnamefont {Martinez}},
  \bibinfo {author} {\bibfnamefont {S.}~\bibnamefont {Oughton}}, \ and\
  \bibinfo {author} {\bibfnamefont {D.}~\bibnamefont {Montgomery}},\
  }\href@noop {} {\bibfield  {journal} {\bibinfo  {journal} {Physica D}\
  }\textbf {\bibinfo {volume} {51}},\ \bibinfo {pages} {531} (\bibinfo {year}
  {1991})}\BibitemShut {NoStop}%
\bibitem [{\citenamefont {Montgomery}\ \emph {et~al.}(1992)\citenamefont
  {Montgomery}, \citenamefont {Shan},\ and\ \citenamefont
  {Matthaeus}}]{Montgomery1993}%
  \BibitemOpen
  \bibfield  {author} {\bibinfo {author} {\bibfnamefont {D.}~\bibnamefont
  {Montgomery}}, \bibinfo {author} {\bibfnamefont {X.}~\bibnamefont {Shan}}, \
  and\ \bibinfo {author} {\bibfnamefont {W.~H.}\ \bibnamefont {Matthaeus}},\
  }\href@noop {} {\bibfield  {journal} {\bibinfo  {journal} {Phys. Fluids}\
  }\textbf {\bibinfo {volume} {A4}},\ \bibinfo {pages} {3} (\bibinfo {year}
  {1992})}\BibitemShut {NoStop}%
\bibitem [{\citenamefont {Li}\ and\ \citenamefont {Montgomery}(1996)}]{Li1996}%
  \BibitemOpen
  \bibfield  {author} {\bibinfo {author} {\bibfnamefont {S.}~\bibnamefont
  {Li}}\ and\ \bibinfo {author} {\bibfnamefont {D.}~\bibnamefont
  {Montgomery}},\ }\href@noop {} {\bibfield  {journal} {\bibinfo  {journal}
  {Phys. Lett. A}\ }\textbf {\bibinfo {volume} {218}},\ \bibinfo {pages} {281}
  (\bibinfo {year} {1996})}\BibitemShut {NoStop}%
\bibitem {Note01}
One usually states the problem the other way round. Given 
the macroscopic Euler equation (\ref{eqn:macroeuler}), which is the physical 
equation coming from fluid mechanics, it can be proven \cite{Mp} that a continuous 
vorticity field $\omega({\vr},t)$ may be approximated arbitrarily well over a 
finite time interval by a collection of point vortices with circulation 
$\Omega\sim 1/N$ as $N\rightarrow +\infty$.
\bibitem [{\citenamefont {Chavanis}(2001)}]{Chavanis2001}%
  \BibitemOpen
  \bibfield  {author} {\bibinfo {author} {\bibfnamefont {P.~H.}\ \bibnamefont
  {Chavanis}},\ }\href@noop {} {\bibfield  {journal} {\bibinfo  {journal}
  {Phys. Rev. E}\ }\textbf {\bibinfo {volume} {64}},\ \bibinfo {pages} {026309}
  (\bibinfo {year} {2001})}\BibitemShut {NoStop}%
\bibitem [{\citenamefont {Chavanis}(2008)}]{Chavanis2008}%
  \BibitemOpen
  \bibfield  {author} {\bibinfo {author} {\bibfnamefont {P.~H.}\ \bibnamefont
  {Chavanis}},\ }\href@noop {} {\bibfield  {journal} {\bibinfo  {journal}
  {Physica A}\ }\textbf {\bibinfo {volume} {387}},\ \bibinfo {pages} {1123}
  (\bibinfo {year} {2008})}\BibitemShut {NoStop}%
\bibitem {Note02}
A limitation of this equation is that it neglects some collective effects. 
In the case of axisymmetric flows, collective effects have been taken into 
account by Dubin and O'Neil \cite{DubinONeil1988,Dubin2003} and Chavanis 
\cite{Chavanis2012,Chavanis2012b}.
\bibitem [{\citenamefont {Chavanis}\ and\ \citenamefont {Lemou}(2007)}]{cl}%
  \BibitemOpen
  \bibfield  {author} {\bibinfo {author} {\bibfnamefont {P.~H.}\ \bibnamefont
  {Chavanis}}\ and\ \bibinfo {author} {\bibfnamefont {M.}~\bibnamefont
  {Lemou}},\ }\href@noop {} {\bibfield  {journal} {\bibinfo  {journal} {Eur.
  Phys. J. B}\ }\textbf {\bibinfo {volume} {59}},\ \bibinfo {pages} {217}
  (\bibinfo {year} {2007})}\BibitemShut {NoStop}%
\bibitem [{\citenamefont {Chavanis}(2012{\natexlab{a}})}]{Chavanis2012}%
  \BibitemOpen
  \bibfield  {author} {\bibinfo {author} {\bibfnamefont {P.~H.}\ \bibnamefont
  {Chavanis}},\ }\href@noop {} {\bibfield  {journal} {\bibinfo  {journal} {J.
  Stat. Mech.}\ }\textbf {\bibinfo {volume} {02}},\ \bibinfo {pages} {P02019}
  (\bibinfo {year} {2012}{\natexlab{a}})}\BibitemShut {NoStop}%
\bibitem [{\citenamefont {Chavanis}(2012{\natexlab{b}})}]{Chavanis2012b}%
  \BibitemOpen
  \bibfield  {author} {\bibinfo {author} {\bibfnamefont {P.~H.}\ \bibnamefont
  {Chavanis}},\ }\href@noop {} {\bibfield  {journal} {\bibinfo  {journal}
  {Physica A}\ }\textbf {\bibinfo {volume} {391}},\ \bibinfo {pages} {3657}
  (\bibinfo {year} {2012}{\natexlab{b}})}\BibitemShut {NoStop}%
\bibitem [{\citenamefont {Chavanis}(2010)}]{Chavanis2010}%
  \BibitemOpen
  \bibfield  {author} {\bibinfo {author} {\bibfnamefont {P.~H.}\ \bibnamefont
  {Chavanis}},\ }\href@noop {} {\bibfield  {journal} {\bibinfo  {journal} {J.
  Stat. Mech.}\ }\textbf {\bibinfo {volume} {05}},\ \bibinfo {pages} {P05019}
  (\bibinfo {year} {2010})}\BibitemShut {NoStop}%
\bibitem [{\citenamefont {Heyvaerts}(2010)}]{Heyvaerts}%
  \BibitemOpen
  \bibfield  {author} {\bibinfo {author} {\bibfnamefont {J.}~\bibnamefont
  {Heyvaerts}},\ }\href@noop {} {\bibfield  {journal} {\bibinfo  {journal}
  {Month. Not. Royal Astron. Soc.}\ }\textbf {\bibinfo {volume} {407}},\
  \bibinfo {pages} {355} (\bibinfo {year} {2010})}\BibitemShut {NoStop}%
\bibitem [{\citenamefont {Chavanis}(2012{\natexlab{c}})}]{Chavanis2012c}%
  \BibitemOpen
  \bibfield  {author} {\bibinfo {author} {\bibfnamefont {P.~H.}\ \bibnamefont
  {Chavanis}},\ }\href@noop {} {\bibfield  {journal} {\bibinfo  {journal}
  {Physica A}\ }\textbf {\bibinfo {volume} {391}},\ \bibinfo {pages} {3680}
  (\bibinfo {year} {2012}{\natexlab{c}})}\BibitemShut {NoStop}%
\bibitem [{\citenamefont {Chavanis}(2013)}]{Chavanis2013}%
  \BibitemOpen
  \bibfield  {author} {\bibinfo {author} {\bibfnamefont {P.~H.}\ \bibnamefont
  {Chavanis}},\ }\href@noop {} {\bibfield  {journal} {\bibinfo  {journal}
  {Astron. Astrophys.}\ }\textbf {\bibinfo {volume} {556}},\ \bibinfo {pages}
  {A93} (\bibinfo {year} {2013})}\BibitemShut {NoStop}%
\bibitem [{\citenamefont {Kawahara}\ and\ \citenamefont
  {Nakanishi}(2007)}]{kn}%
  \BibitemOpen
  \bibfield  {author} {\bibinfo {author} {\bibfnamefont {R.}~\bibnamefont
  {Kawahara}}\ and\ \bibinfo {author} {\bibfnamefont {H.}~\bibnamefont
  {Nakanishi}},\ }\href@noop {} {\bibfield  {journal} {\bibinfo  {journal} {J.
  Phys. Soc. Jpn.}\ }\textbf {\bibinfo {volume} {76}},\ \bibinfo {pages}
  {074001} (\bibinfo {year} {2007})}\BibitemShut {NoStop}%
\bibitem [{\citenamefont {Chavanis}(1998)}]{chavanis1998}%
  \BibitemOpen
  \bibfield  {author} {\bibinfo {author} {\bibfnamefont {P.~H.}\ \bibnamefont
  {Chavanis}},\ }\href@noop {} {\bibfield  {journal} {\bibinfo  {journal}
  {Phys. Rev. E}\ }\textbf {\bibinfo {volume} {58}},\ \bibinfo {pages} {R1199}
  (\bibinfo {year} {1998})}\BibitemShut {NoStop}%
\bibitem [{\citenamefont {Dubin}(2003)}]{Dubin2003}%
  \BibitemOpen
  \bibfield  {author} {\bibinfo {author} {\bibfnamefont {D.~H.~E.}\
  \bibnamefont {Dubin}},\ }\href@noop {} {\bibfield  {journal} {\bibinfo
  {journal} {Phys. Plasmas}\ }\textbf {\bibinfo {volume} {10}},\ \bibinfo
  {pages} {1338} (\bibinfo {year} {2003})}\BibitemShut {NoStop}%
\bibitem [{\citenamefont {Dubin}\ and\ \citenamefont
  {O'Neil}(1988)}]{DubinONeil1988}%
  \BibitemOpen
  \bibfield  {author} {\bibinfo {author} {\bibfnamefont {D.~H.~E.}\
  \bibnamefont {Dubin}}\ and\ \bibinfo {author} {\bibfnamefont {T.~M.}\
  \bibnamefont {O'Neil}},\ }\href@noop {} {\bibfield  {journal} {\bibinfo
  {journal} {Phys. Rev. Lett.}\ }\textbf {\bibinfo {volume} {60}},\ \bibinfo
  {pages} {1286} (\bibinfo {year} {1988})}\BibitemShut {NoStop}%
\bibitem [{\citenamefont {Taylor}\ and\ \citenamefont
  {McNamara}(1971)}]{TaylorMcNamara1971}%
  \BibitemOpen
  \bibfield  {author} {\bibinfo {author} {\bibfnamefont {J.~B.}\ \bibnamefont
  {Taylor}}\ and\ \bibinfo {author} {\bibfnamefont {B.}~\bibnamefont
  {McNamara}},\ }\href@noop {} {\bibfield  {journal} {\bibinfo  {journal}
  {Phys. Fluids}\ }\textbf {\bibinfo {volume} {14}},\ \bibinfo {pages} {1492}
  (\bibinfo {year} {1971})}\BibitemShut {NoStop}%
\bibitem [{\citenamefont {Dawson}\ \emph {et~al.}(1971)\citenamefont {Dawson},
  \citenamefont {Okuda},\ and\ \citenamefont
  {Carlile}}]{DawsonOkudaCarlile1971}%
  \BibitemOpen
  \bibfield  {author} {\bibinfo {author} {\bibfnamefont {J.~M.}\ \bibnamefont
  {Dawson}}, \bibinfo {author} {\bibfnamefont {H.}~\bibnamefont {Okuda}}, \
  and\ \bibinfo {author} {\bibfnamefont {R.~N.}\ \bibnamefont {Carlile}},\
  }\href@noop {} {\bibfield  {journal} {\bibinfo  {journal} {Phys. Rev. Lett.}\
  }\textbf {\bibinfo {volume} {27}},\ \bibinfo {pages} {491} (\bibinfo {year}
  {1971})}\BibitemShut {NoStop}%
\bibitem [{\citenamefont {Okuda}\ and\ \citenamefont
  {Dawson}(1973)}]{OkudaDawson1973}%
  \BibitemOpen
  \bibfield  {author} {\bibinfo {author} {\bibfnamefont {H.}~\bibnamefont
  {Okuda}}\ and\ \bibinfo {author} {\bibfnamefont {J.~M.}\ \bibnamefont
  {Dawson}},\ }\href@noop {} {\bibfield  {journal} {\bibinfo  {journal} {Phys.
  Fluids}\ }\textbf {\bibinfo {volume} {16}},\ \bibinfo {pages} {408} (\bibinfo
  {year} {1973})}\BibitemShut {NoStop}%
\bibitem {Note03}
For example, this approach could be suitable to analyze the breaking of 
``vortex crystals'' \cite{Fine1995,kn} due to finite $N$ effects (collisions).
\bibitem [{\citenamefont {Klimontovich}(1967)}]{Klimontovich}%
  \BibitemOpen
  \bibfield  {author} {\bibinfo {author} {\bibfnamefont {Y.~L.}\ \bibnamefont
  {Klimontovich}},\ }\href@noop {} {\emph {\bibinfo {title} {The statistical
  theory of non-equilibrium processes in a plasma}}}\ (\bibinfo  {publisher}
  {MIT Press},\ \bibinfo {address} {Cambridge, Massachusetts},\ \bibinfo {year}
  {1967})\BibitemShut {NoStop}%
\bibitem {Note04}
Actually the form of this collision kernel was ``guessed'' by one of us (PHC) a 
long time ago, in order to conserve the energy, before finding that the 
conservation of energy could also be achieved by the delta function in Eq. 
(\ref{grd}).
\bibitem [{\citenamefont {Fine}\ \emph {et~al.}(1995)\citenamefont {Fine},
  \citenamefont {Cass}, \citenamefont {Flynn},\ and\ \citenamefont
  {Driscoll}}]{Fine1995}%
  \BibitemOpen
  \bibfield  {author} {\bibinfo {author} {\bibfnamefont {K.~S.}\ \bibnamefont
  {Fine}}, \bibinfo {author} {\bibfnamefont {A.~C.}\ \bibnamefont {Cass}},
  \bibinfo {author} {\bibfnamefont {W.~G.}\ \bibnamefont {Flynn}}, \ and\
  \bibinfo {author} {\bibfnamefont {C.~F.}\ \bibnamefont {Driscoll}},\
  }\href@noop {} {\bibfield  {journal} {\bibinfo  {journal} {Phys. Rev. Lett.}\
  }\textbf {\bibinfo {volume} {75}},\ \bibinfo {pages} {3277} (\bibinfo {year}
  {1995})}\BibitemShut {NoStop}%
\bibitem [{Not({\natexlab{e}})}]{Note05}
The collisional evolution may also be described by the kinetic equation 
(\ref{grd}) that approaches the Boltzmann distribution without necessarily 
reaching it (see the discussion in \cite{Chavanis2010}).
\bibitem [{\citenamefont {Marchioro}\ and\ \citenamefont
  {Pulvirenti}(1994)}]{Mp}%
  \BibitemOpen
  \bibfield  {author} {\bibinfo {author} {\bibfnamefont {C.}~\bibnamefont
  {Marchioro}}\ and\ \bibinfo {author} {\bibfnamefont {M.}~\bibnamefont
  {Pulvirenti}},\ }\href@noop {} {\emph {\bibinfo {title} {Mathematical Theory
  of Incompressible Nonviscous Fluids}}}\ (\bibinfo  {publisher} {Springer},\
  \bibinfo {address} {New-York},\ \bibinfo {year} {1994})\BibitemShut {NoStop}%
\end{thebibliography}
%


%

\end{document}